\documentclass[aps,prb,nobibnotes,twocolumn,superscriptaddress,bibliography]{revtex4-2}
\usepackage{amsfonts}
\usepackage{mathrsfs}
\usepackage{amsmath}
\usepackage{color}
\usepackage{graphicx}
\usepackage{bm}
\usepackage{amssymb}
\usepackage{xspace}
\usepackage{epstopdf}
\usepackage{dcolumn}
\usepackage{longtable}
\usepackage{multirow}
\usepackage{float}
\usepackage{comment}

\usepackage[colorlinks=true, letterpaper=true, pdfstartview=FitV, linkcolor=blue, citecolor=blue, urlcolor=blue]{hyperref}

\makeatother

\begin{document}
\title{Anomalous shift in scattering from topological nodal-ring semimetals}
\author{Runze Li}
\affiliation{Research Laboratory for Quantum Materials, Department of Applied Physics, The Hong Kong Polytechnic University, Kowloon, Hong Kong, China}
\affiliation{Key Lab of Advanced Optoelectronic Quantum Architecture and Measurement (MOE), Beijing Key Lab of Nanophotonics \& Ultrafine Optoelectronic Systems, and School of Physics, Beijing Institute of Technology, Beijing 100081, China}

\author{Chaoxi Cui}
\affiliation{Key Lab of Advanced Optoelectronic Quantum Architecture and Measurement (MOE), Beijing Key Lab of Nanophotonics \& Ultrafine Optoelectronic Systems, and School of Physics, Beijing Institute of Technology, Beijing 100081, China}

\author{Ying Liu}
\affiliation{School of Materials Science and Engineering, Hebei University of Technology, Tianjin 300130, China}

\author{Zhi-Ming Yu}
\email{zhiming\_yu@bit.edu.cn}
\affiliation{Key Lab of Advanced Optoelectronic Quantum Architecture and Measurement (MOE), Beijing Key Lab of Nanophotonics \& Ultrafine Optoelectronic Systems, and School of Physics, Beijing Institute of Technology, Beijing 100081, China}
\affiliation{Beijing Institute of Technology, Zhuhai 519088, China}

\author{Shengyuan A. Yang}
\email{shengyuan.yang@polyu.edu.hk}
\affiliation{Research Laboratory for Quantum Materials, Department of Applied Physics, The Hong Kong Polytechnic University, Kowloon, Hong Kong, China}

\begin{abstract}
An electron beam may experience an anomalous spatial shift during an interface scattering process. Here, 
we investigate this phenomenon for reflection from mirror-symmetry-protected nodal-ring semimetals, which 
are characterized by an integer topological charge $\chi_h$. We show that the shift is generally enhanced by the presence of nodal rings, and the ring's geometry can be inferred from the
profile of shift vectors in the interface momentum plane. Importantly, the anomalous shift encodes the 
topological information of the ring, where the circulation of the shift vector field $\kappa_s$ over a semicircle is governed by the topological charge, with a simple relationship: $\kappa_s=-2\pi \chi_h$. Furthermore, we demonstrate that the shift and its circulation reflect distinct features of topological phase transitions of the charged rings. This study uncovers a novel physical signature of topological nodal rings and positions anomalous scattering shifts as a powerful tool for probing topological band structures.
\end{abstract}
\maketitle

\section{Introduction}
Topological semimetals have been attracting tremendous interest in condensed matter physics research~\cite{Chiu2016,Bansil2016,Yang2016,Armitage2018,LvBQ2021,yu2022}. Unlike conventional metals,  they have special band crossings near Fermi level, which are  stabilized by symmetry and/or topology. In a three dimensional (3D) material, the possible band crossings can be
categorized as 0D nodal points~\cite{Wan2011,Shuichi2007,weyl2,weyl3,Young2012,Wang2012}, 1D nodal lines~\cite{YangSA2014,Mullen2015,chenY2015,YuR2015,Kim2015,FangC2015}, and 2D nodal surfaces~\cite{zhong2016,Liang2016,Wu2018}. The prime example of a nodal-point semimetal is the Weyl semimetals~\cite{Wan2011,Shuichi2007,weyl2,weyl3}, in which two bands cross at isolated nodal points, known as Weyl points, at Fermi level. Each Weyl point is characterized by a quantized nontrivial topological charge, corresponding to the Chern number defined on a 2D sphere that encloses the Weyl point.

As mentioned, two bands crossing at Fermi level may also form 1D nodal lines~\cite{YangSA2014,Mullen2015,chenY2015,YuR2015,Kim2015,FangC2015}. Such lines have to be closed in momentum space, so they are often referred to also as nodal loops or nodal rings.
A stable nodal line requires certain symmetry protection. For example, in spinless systems (i.e., with negligible spin-orbit coupling), a nodal line can be protected by the combined spacetime inversion ($PT$) symmetry~\cite{Mullen2015,chenY2015,YuR2015}.
Another more commonly encountered case is with protection by mirror symmetry~\cite{YangSA2014,FangC2015}. For instance, the mirror symmetry $M_z$
can protect a nodal ring in a mirror-invariant plane, i.e.,
the $k_z=0$ or $k_z=\pi$ plane, if the two crossing bands have opposite mirror eigenvalues in this plane.
In the simplest case, a mirror-protected nodal ring does not have a topological charge like that of a Weyl point. And
under strong enough variations, e.g., by pulling the two bands apart in energy, the ring may shrink to a point and get completely removed, leaving a fully gapped system~\cite{chenY2015,LiSi2018}.
Interestingly, later research reveals that for mirror-protected nodal rings, one may define
a topological charge $\chi_h$ on a mirror-symmetric hemisphere~\cite{nl1,nl2}. For nontrivial $\chi_h\neq 0$, the nodal ring, also called vortex ring~\cite{nl1}, cannot be fully gapped but may evolve into a pair of Weyl points after it shrinks to a point.

A current focus of research on topological semimetals is the exploration of physical signatures associated with topological nodal structures. In particular, the potential manifestations of topological charges, such as $\chi$ for the mirror-protected nodal rings, remain underexplored, presenting an important challenge for ongoing investigation.

When a light beam is reflected from a material interface, the outgoing beam may undergo a spatial shift from the incident beam on the interface plane. This shift has a longitudinal component and a transverse component with respect to the plane of incidence, and they are referred to as Goos-H\"{a}nchen shift~\cite{Goos1947} and Imbert-Fedorov shift~\cite{IF1955,IF1972}, respectively.
These optical phenomena are well known and have been extensively studied before~\cite{Bliokh_2013,bliokh2015}. There exists analogous effect when the light beam is replaced by an electron beam~\cite{Miller1972,Fra1974,Sinitsyn2005,Chen2008,Beenakker2009,ZW2011,chen2011goos,Jiang2015,Yang2015,Yu2019,Ch2019,Do2022}. Interestingly, previous studies of such anomalous electronic shift in reflection from systems such as
graphene~\cite{Beenakker2009}, Weyl semimetal~\cite{Jiang2015,Yang2015}, and nodal superconductors~\cite{Yliu2017,Yu2018,1Yliu2018,2Yliu2018,Li2024} revealed that the effect is particularly sensitive to the phase winding of electron wave function. Since topological nodal structures can be viewed as singularities for the phase winding in momentum space, it is natural to anticipate that they should strongly influence the shift. Indeed, strong effects have been predicted in graphene and Weyl semimetal (which host nodal points), leading to modified waveguide mode~\cite{Beenakker2009,1Yliu2018} and chirality-dependent Hall effect~\cite{Yang2015}.
Furthermore, recent studies unveil that the topological charge of a Weyl point can make a quantized fingerprint in the anomalous scattering shift. Specifically, the circulation of the anomalous shift (CAS) $\kappa$ along a closed path in the interface momentum plane assumes quantized value $\kappa=2\pi N$, where $N$ just corresponds to the net charge of Weyl points enclosed by the path~\cite{Yliu2020}. This correspondence gives a remarkable manifestation of topological charge in physical effects, and allows its value to be
directly probed from anomalous shift. Motivated by these advances, a natural question is: What will be the impact of nodal rings on anomalous scattering shift? Particularly, does the charge $\chi_h$ leave a fingerprint in this effect?

In this work, we theoretically address these questions by systematically investigating the anomalous reflection shift in topological semimetals with mirror-protected nodal rings. We demonstrate that nodal rings, regardless of whether they carry a trivial or nontrivial topological charge $\chi_h$, can significantly enhance the shift. Consequently, both the range and shape of the ring can be inferred by mapping the shift profile. Remarkably, we find that the correspondence between CAS and topological charge for Weyl points can be generalized to nodal rings, subject to certain refinements and symmetry constraints. Specifically, we show that it is the CAS $\kappa_s$ defined over a semicircle that captures the topological charge $\chi_h$, with the relationship:
\begin{equation}\label{rel}
  \kappa_s=-2\pi \chi_h.
\end{equation}
Therefore, measuring the anomalous shift enables the direct detection of the topological charge of vortex rings. Additional features of nodal rings and their associated topological phase transitions that are manifested in the shift are also discussed. These findings reveal a novel signature of nodal-ring semimetals and establish anomalous scattering shifts as a powerful probe of topological band structures.

\section{nodal ring model}\label{secII}
In this section, we shall construct a model for mirror-protected nodal rings. As an advantage, depending on its parameters, this model captures both the conventional nodal rings (i.e., with $\chi_h=0$) as well as the vortex rings(i.e., with nontrivial $\chi_h$)~\cite{nl1,nl2}. In addition, previous works mainly studied models with $|\chi_h|=1$. Our construction here realizes $\chi_h$ with arbitrary integer values.

Consider a 3D system described by Hamiltonian $H(\boldsymbol{k})$, which preserves mirror reflection symmetry $M_z$. The symmetry constraint is
\begin{equation}
M_z H\left(k_x, k_y,-k_z\right) M_z^{-1}=H\left(k_x, k_y, k_z\right).
\end{equation}
The two high-symmetry planes $k_z=0$ and $k_z=\pi$ of the Brillouin zone are the $M_z$-invariant planes. On these two planes, each band has a definite $M_z$ eigenvalue, and two bands with opposite $M_z$ eigenvalues can cross to form a protected nodal ring. Without loss of generality, we shall assume the considered ring is in the $k_z=0$ plane in the following discussion.

Before presenting the model, we first introduce the definition of the topological charge $\chi_h$ for mirror-protected band degeneracies. Recall that the topological charge $N$ of a Weyl point is defined by the Chern number on a closed surface $S$ enclosing the point. Usually, we take $S$ to be a sphere, and the Chern number can be evaluated by the integral of Berry curvature over $S$, i.e.,
\begin{equation}
  N=\frac{1}{2\pi}\oint_S \bm\Omega \cdot d\bm\sigma,
\end{equation}
where $\bm\Omega$ is the total Berry curvature for all valence bands.

\begin{figure}[t]
	\includegraphics[width=8.7cm]{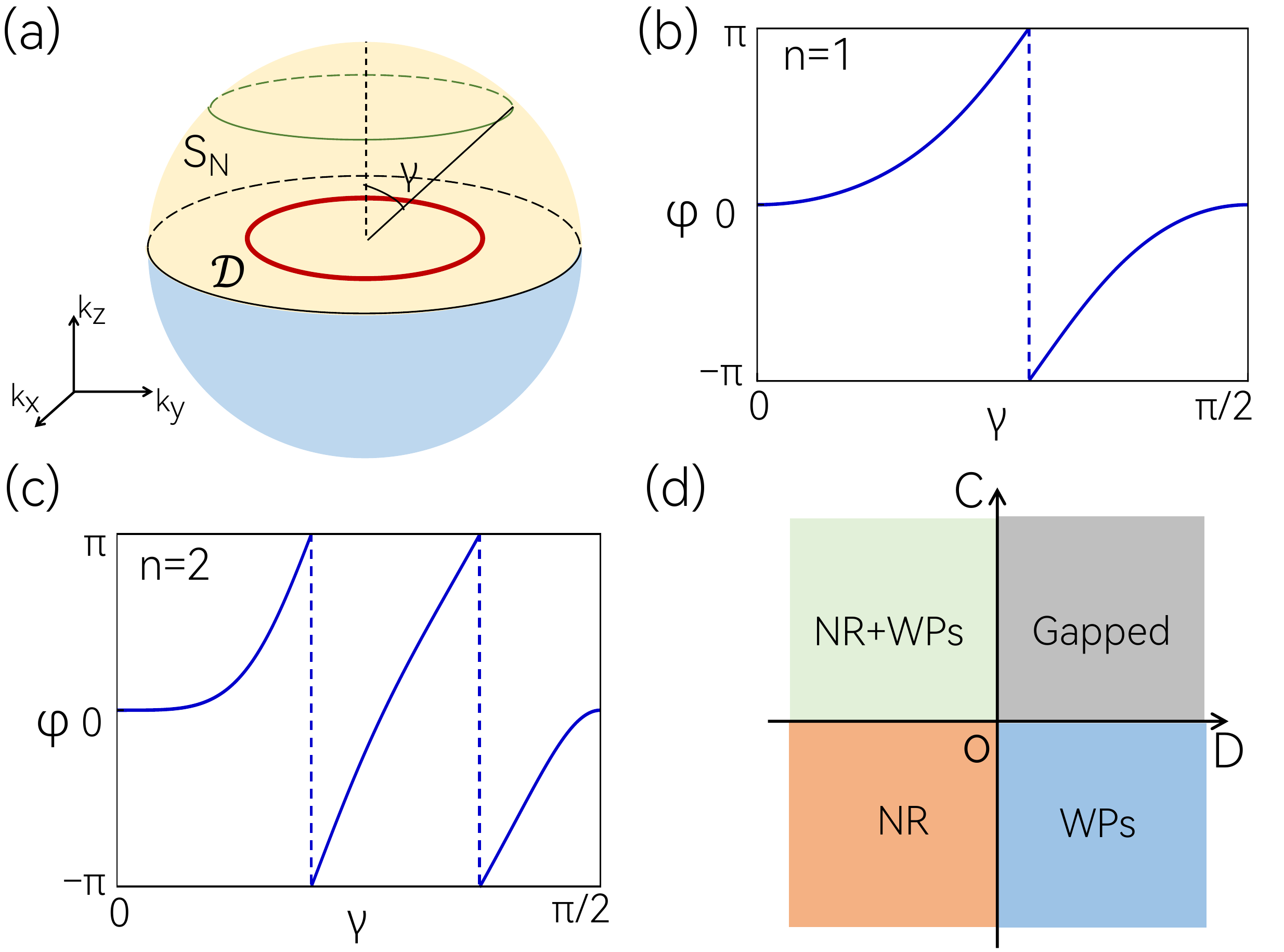}
	\caption{(a) Illustration of a mirror-protected nodal ring, denoted by the red circle in $k_z=0$ plane. The ring is inside the disk $\mathcal{D}$ and covered by hemisphere $S_N$.  (b) and (c) show the Wilson loop on a circle of latitude versus the spherical angle $\gamma$ (as shown in (1)) for model Eq.~(\ref{HH}) with $n=1$  and $n=2$, respectively. Here, we set $A=1$, $B=1$, $C=-0.2$ and $D=-2$. (d) Phase diagram of  model (\ref{HH}) in the $C$-$D$ parameter plane. Here, we take $B>0$.
		\label{nodalstructure}}
\end{figure}

Now, if the system has $M_z$ symmetry and some
$M_z$ protected nodal ring in the $k_z=0$ plane, we may have a refined topological classification enabled by the mirror symmetry. Consider the closed surface illustrated in Fig.~\ref{nodalstructure}(a), which consists of two parts: the hemisphere $S_N$ and the disk $\mathcal{D}$ containing the ring in the $k_z=0$ plane. The Berry curvature field is constrained in the mirror plane. Since $\bm\Omega$ is a pseudovector, we must have $\Omega_x=\Omega_y=0$ in the mirror plane. As for $\Omega_z$, it must also vanish since the two bands that form the nodal ring have opposite $M_z$ eigenvalue and are decoupled in $\mathcal{D}$~\cite{Lilei2023}.
This result is exact for a two-band model. It does not hold when more bands are involved. Nevertheless,
if the nodal ring is well separated from other irrelevant bands in energy, this result should still hold as a good approximation. Then, because of the vanishing Berry curvature field on $\mathcal{D}$, one can see that
\begin{equation}\label{chi}
  \chi_h=\frac{1}{2\pi}\int_{S_N} \bm\Omega \cdot d\bm\sigma
\end{equation}
naturally defines an integer valued topological charge for the mirror-protected nodal ring~\cite{nl2}. In the above discussion, we take $S_N$ to be a hemisphere, but one can easily see that its shape is flexible, as long as its boundary encloses the ring in the mirror plane. This charge $\chi_h$ gives a refined characterization compared to $N$. This can be seen by noting that  due to the mirror symmetry, the total $N$ over the spherical surface in Fig.~\ref{nodalstructure}(a) must vanish, however, $\chi_h$ over the northern and southern hemispheres have opposite signs and may not be zero.

Next, we present our model for the mirror-protected nodal ring in $k_z=0$ plane:
\begin{equation}\label{HH}
	H=Ak_z (k_-^n \sigma_{+}+k_+^n \sigma_{-})+\Big[B(k_x^2+k_y^2)+C k_z^2+D\Big] \sigma_z,
\end{equation}
where energy is measured from the nodal ring,
$k_{\pm}\equiv k_x \pm i k_y$, $\sigma_{ \pm}\equiv\left(\sigma_x \pm i \sigma_y\right) / 2$ with the Pauli matrices denoting the two band degree of freedom, $A$, $B$, $C$ and $D$ are real model parameters.  This model actually describes several interesting topological states.
The nodal ring exists when the parameters
$B$ and $D$ have opposite signs, i.e., $BD<0$. In such a case, a ring with radius
\begin{equation}\label{RR}
  R = \sqrt{-D/B}
\end{equation}
appears in the $k_z = 0$ plane. Without loss of generality, let's assume $B>0$ in the following. Then the ring exists for $D<0$.

The topological charge of  the nodal ring depend on
$n$. For the case with $n=0$, it reduces to the conventional mirror-protected nodal ring (under $B>0$, $D<0$), with
\begin{equation}\label{model1}
	H=Ak_z \sigma_{x}+\Big[B(k_x^2+k_y^2)+C k_z^2+D\Big] \sigma_z.
\end{equation}
It has $\chi_h=0$. This can be easily seen by noting that model (\ref{model1}) preserves both
inversion $P=\sigma_z$ and time reversal $T=\sigma_z \mathcal{K}$ symmetries ($\mathcal K$ is complex conjugation),  so Berry curvature must vanish and so does $\chi_h$.

\begin{figure}[t]
\includegraphics[width=8.7cm]{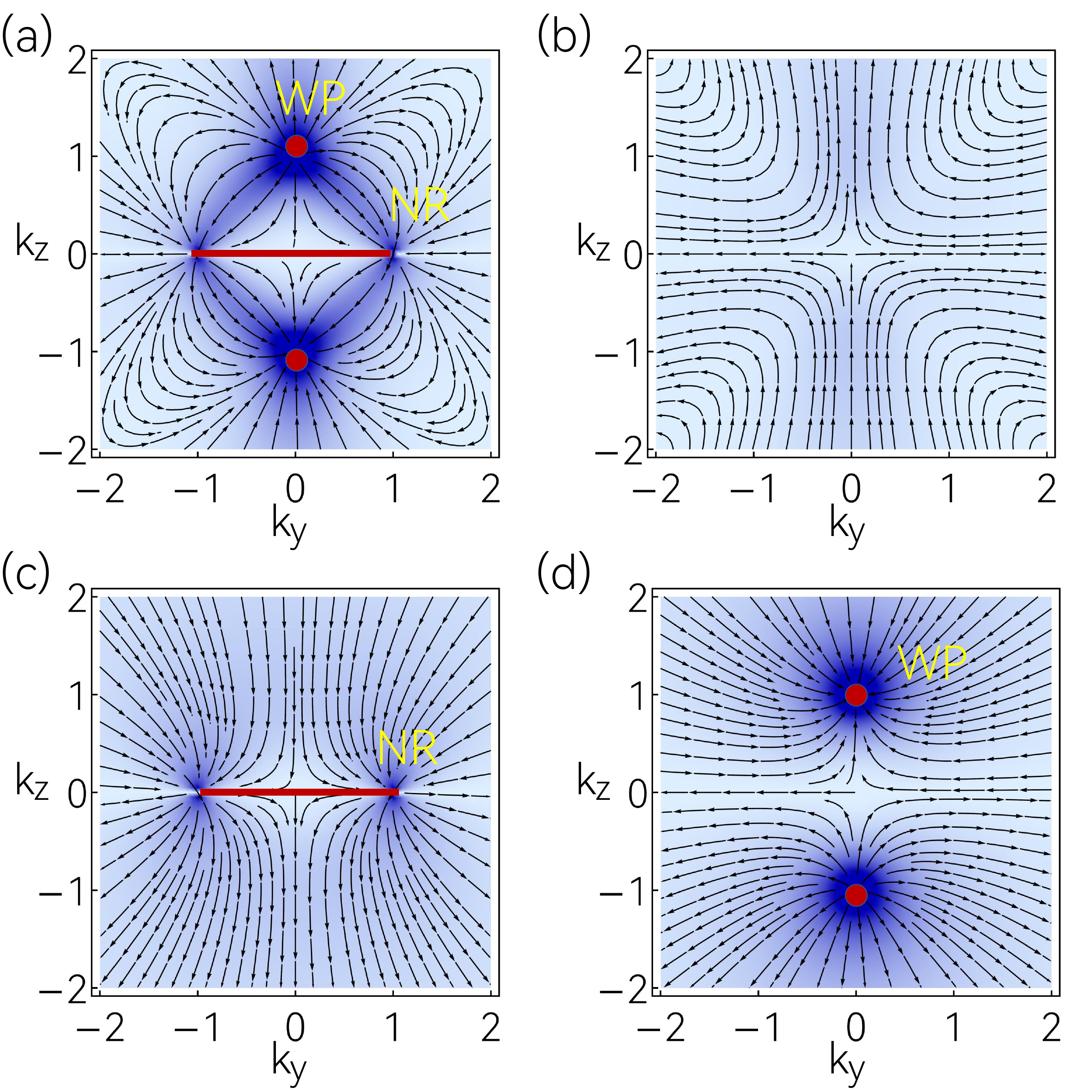}
\caption{Distribution of the Berry curvature field $\boldsymbol{\Omega}(\boldsymbol{k})$ for the four phases in the phase diagram Fig.~\ref{nodalstructure}(d). The plot is for the $k_x=0$ plane. The color map indicates the magnitude of Berry curvature. (a) The phase with nodal ring plus Weyl points, for $C>0$ and $D<0$ . (b) The gapped phase, for $C>0$ and $D>0$. (c) The phase with nodal ring, for $C<0$ and $D<0$. (d) The phase with Weyl points, for $C<0$ and $D>0$. In the calculation, we set $n=1$, $A=1$, $B=0.5$, $|C|=0.5$ and $|D|=0.5$.
	\label{BerryC}}
\end{figure}

Vortex ring having nontrivial $\chi_h$ appears when $n\geq 1$. In addition to $B>0$, $D<0$, we take $C<0$, then
the model (\ref{HH}) contains the nodal ring as the only band degeneracy. In this case, we have
\begin{equation}\label{chihn}
  \chi_h=-n.
\end{equation}
This can be shown by directly computing the integral in Eq.~(\ref{chi}). Another way is via the Wilson loop method, by computing the
evolution of the Berry phase $\varphi(\gamma)$ over a latitude as its angle $\gamma$ varies from $0$ to $\pi/2$ [see Fig.~\ref{nodalstructure}(a)].
In Figs.~\ref{nodalstructure}(b) and ~\ref{nodalstructure}(c), we plot this Berry phase evolution for the cases with $n=1$ and $n=2$, $\chi_h$ is read off from the winding number of $\varphi$ across $2\pi$, which confirms the relation (\ref{chihn}).

In fact, the nodal ring case is only one possible phase of model (\ref{HH}). Keeping $B>0$ and considering the possible values of $C$ and $D$, we can have four phase regions in the $C$-$D$ parameter space, as shown in Fig.~\ref{nodalstructure}(d).
Notably, when $C$ and $D$ have opposite signs, there appear a pair of Weyl points at zero energy, located on the $k_z$ axis, with
\begin{equation}\label{KW}
  k_z=\pm \sqrt{-D/C}.
\end{equation}
The effective Hamiltonian expanded at the Weyl point takes the form of
\begin{equation}
  \mathcal H_{w_{\pm}} = \pm \sqrt{-{D}/{C}} (A q_-^{n} \sigma_+ + A q_+^{n} \sigma_- + C q_z \sigma_z),
\end{equation}
where $\bm q$ is the momentum measured from the Weyl point. The point $\mathcal H_{w_{\pm}}$ has a topological charge of
$\pm\text{sgn}(C)n$. For $n>1$, such a Weyl point is usually called a multi-Weyl point or a charge-$n$ Weyl point.
Weyl points appear in two regions in Fig.~\ref{nodalstructure}(d). For $C<0$ and $D>0$, Weyl points are the only band degeneracies.
For $C>0$ and $D<0$, they coexist with the vortex ring. The Weyl point above the mirror plane (i.e., having $k_z>0$) has
charge $n$, opposite to that of the ring, i.e., if taking a hemisphere that encloses both the ring and the Weyl point, we shall have $\chi_h=0$. Finally, for region $C>0$ and $D>0$, the model is fully gapped with no nodal features, and hence  $\chi_h=0$.

In Fig.~\ref{BerryC}, we plot the distribution of Berry curvature field the $k_x=0$ plane for the four phases. One observes that both the ring and the Weyl points act as source for Berry curvature, manifesting that they carry nontrivial topological charges.

It is interesting to consider the transitions between the phases in Fig.~\ref{nodalstructure}(d). For example, fixing $C<0$, we may consider the transition when tuning $D$ from negative to postive, across zero. When increasing $D$ towards zero from the negative side, according to Eq.~\ref{RR}, the radius of the ring shrinks to zero. However, unlike conventional nodal rings, the spectrum cannot be fully gapped, because of the nontrivial $\chi_h$ charge of the ring. Instead, a pair of Weyl points emerge after $D$ turns positive, which ensures $\chi_h$ is unchanged. In a sense, the vortex ring is transformed into the Weyl points.

Similarly, we may consider the topological phase transition when $C$ is tuned from negative to positive, while keeping $D<0$. Changing the value of $C$ does not affect the shape of the ring. However, right after $C$ turns positive, there emerges a pair of Weyl points at infinity, according to Eq.~(\ref{KW}). Further increasing $C$ moves the two points from infinity towards the origin. If one calculates $\chi_h$ on a hemisphere with finite radius ($>R$), then initially, one should have $\chi_h=n$, but after the Weyl point crosses the hemisphere which brings a $-n$ charge, $\chi_h$ should jump to zero.

\section{Anomalous shift and its circulation} \label{secIII}

In the following, we first review the concept of anomalous scattering shift and the method for its evaluation. Then, we discuss CAS and how its definition can be adapted to study mirror-protected nodal rings.
\begin{figure}[t]
	\includegraphics[width=8.7cm]{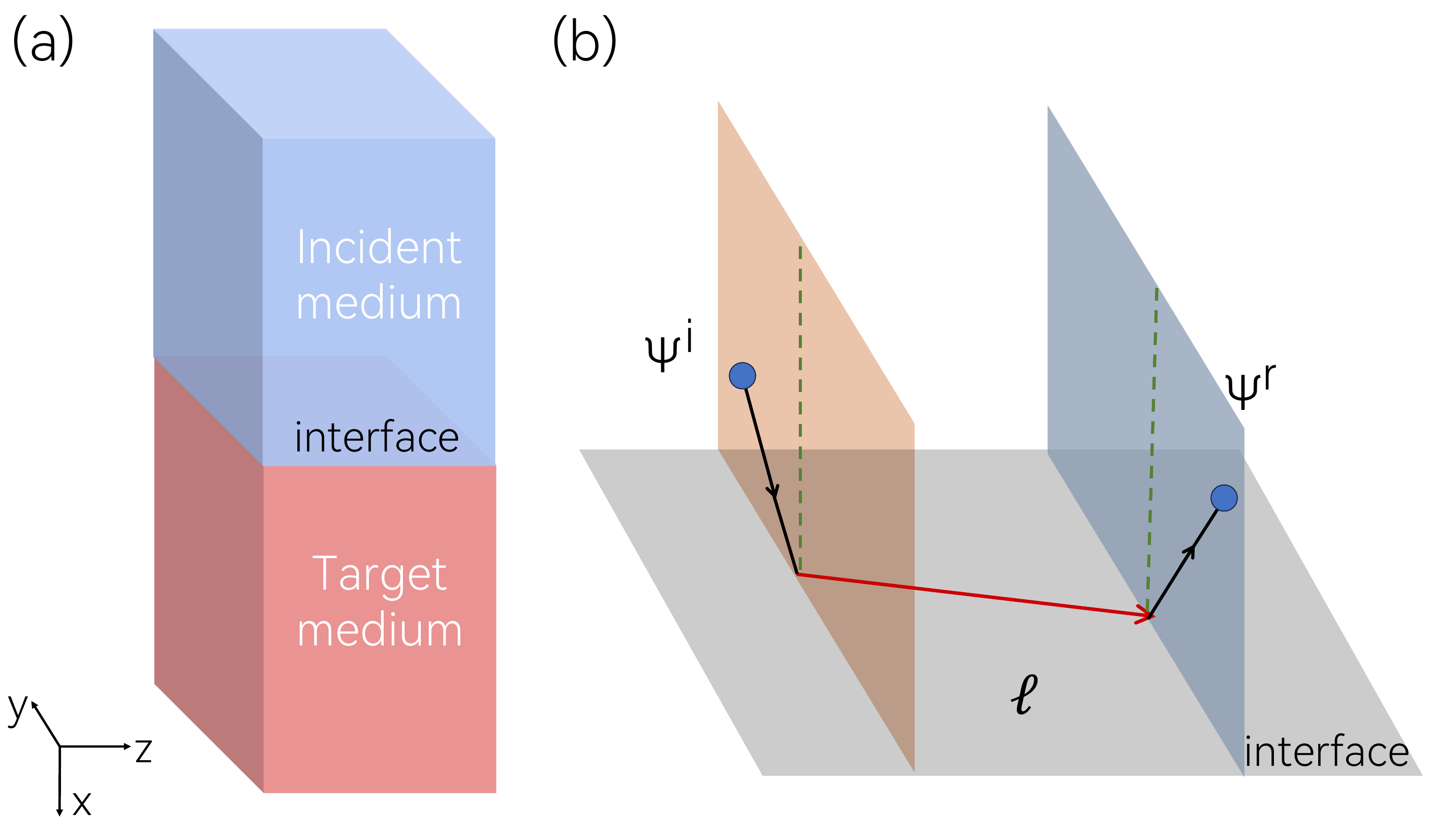}
	\caption{(a) Schematic figure showing the setup. An interface at $x=0$ is formed between an incident medium ($x<0$) and a target medium ($x>0$). (b) An electron beam $\Psi^i$ from the incident medium is scattered at the interface. The reflected beam is described by $\Psi^r$. The reflected beam may experience a spatial shift $\bm \ell$ from the incident point in the interface plane.
		\label{junction}}
\end{figure}

\subsection{Anomalous shift in interface scattering}
Consider a setup with a flat interface between two media at the $x=0$ plane.  A beam of electrons $\Psi^i$ is incoming from the $x<0$ region and is scattered at the interface into a reflected beam $\Psi^r$ and a transmitted beam $\Psi^t$, as illustrated in Fig.~\ref{junction}.

To define the anomalous spatial shift, the beam has to be confined in both real and momentum spaces. Assuming the incident beam has momentum centered at $\boldsymbol{k}^c$, it can generally be expressed as~\cite{Yu2019}
\begin{equation}\label{i beam}
\Psi^i\left(\boldsymbol{r}, \boldsymbol{k}^c\right)=\int  w\left(\boldsymbol{k}-\boldsymbol{k}^c\right) \psi^{i}(\boldsymbol{k})  d \boldsymbol{k},
\end{equation}
where $\psi^{i}(\boldsymbol{k})=e^{i\boldsymbol{k} \cdot \boldsymbol{r}}|u^i(\boldsymbol{k})\rangle$ is the Bloch eigenstate of the incident medium $H_1$, with $|u^i(\boldsymbol{k})\rangle$ being the cell-periodic part, and $w$ is weight distribution of the different partial waves. The specific form of $w$ is not important, as it does not affect the shift. For concreteness, one usually take $w$ to have a Gaussian form.

\begin{figure}[t]
	\includegraphics[width=8.7cm]{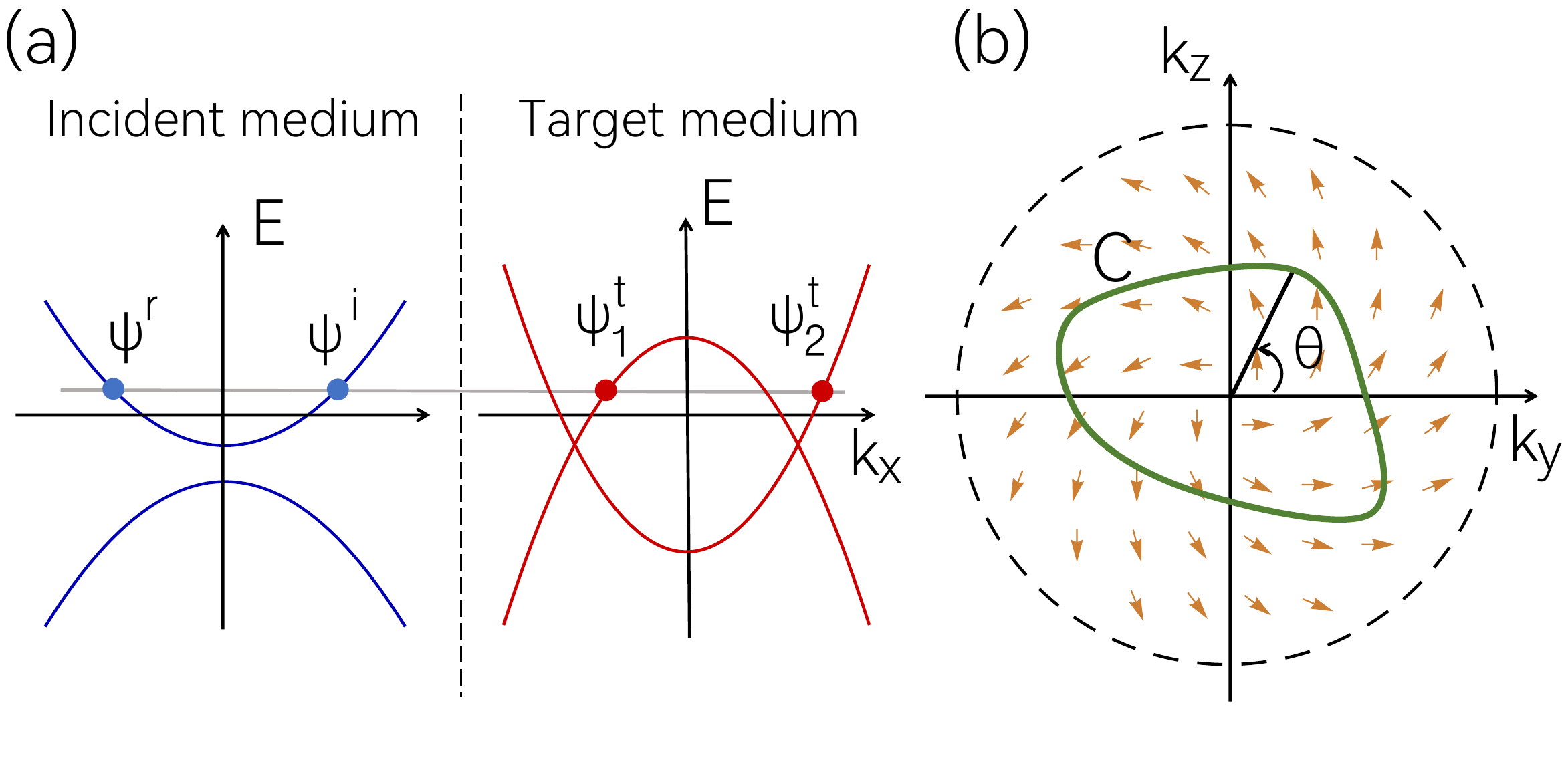}
	\caption{(a) Schematic plot of the band structures of the incident and the target media. The target medium here has a mirror-protected nodal ring near the Fermi level. (b) Illustration of the CAS over a closed loop $C$ in the interface momentum plane.
		\label{band}}
\end{figure}

When the incident beam hits the interface, each partial wave $\psi^{i}(\boldsymbol{k})$ is scattered into reflected waves and transmitted waves, with certain $k$ dependent scattering amplitudes. Here, if the two media involved are each described by a two-band model [see Fig.~\ref{band}(a)], then the scattering states can be readily written down as

\begin{equation}
\psi(\boldsymbol{k})	=	\begin{cases}
\psi^{i}+r_1\psi^{r}_{1}+r_{2}\psi^{r}_{2}, & x<0\\
t_{1}\psi^{t}_{1}+t_{2}\psi^{t}_{2}, & x>0,
\end{cases}
\end{equation}
where $\psi^{r}_{1,2}$ and $\psi^{t}_{1,2}$ represent the reflected and transmitted modes, respectively, and $r$'s and $t$'s are the corresponding amplitudes.
For the model with dispersion shown in Fig.~\ref{band}(a), only one reflected mode is propagating, whereas the other one is an evanescent mode. We shall focus on the propagating one and label it as $\psi^r$ without a subscript.
Then, the reflected beam can be expressed as
\begin{equation}\label{r beam}
\Psi^r\left(\boldsymbol{r}, \boldsymbol{k}^c\right)=\int w\left(\boldsymbol{k}-\boldsymbol{k}^c\right) r(\boldsymbol{k}) \psi^r(\boldsymbol{k}) d \boldsymbol{k}.
\end{equation}

As illustrated in Fig.~\ref{junction}(a), the anomalous shift ${\bm \ell}$ refers to the shift between the two points where the two beams $\Psi^i$ and $\Psi^r$ hit the interface. Clearly, ${\bm \ell}$ is a vector lying in the interface plane, i.e., $y$-$z$ plane here. Furthermore, it is a function of the interface momentum $\bm k_\|$, which is the conserved component of $\bm k^c$ parallel to the interface plane. In fact, previous studies have established that~\cite{Sinitsyn2006,shi2019,Yliu2020}
\begin{equation}
\boldsymbol{\ell}=\boldsymbol{\mathcal A}_r-\boldsymbol{\mathcal A}_i-\frac{\partial \phi}{\partial \boldsymbol{k}_\|},
\end{equation}
where $\phi=\text{arg}(r)$ is the phase angle of reflection amplitude $r$, and $\boldsymbol{\mathcal A}_{i(r)}=\left\langle u^{i(r)}\left|i\partial_{\boldsymbol{k}_{\|}}\right|u^{i(r)}\right\rangle $  is the Berry connection for the corresponding state in the interface momentum plane.

\subsection{CAS on a closed loop}
As shown in Fig.~\ref{band}(b), since $\boldsymbol{\ell}$ forms a vector field over the interface momentum $\boldsymbol{k}_{\|}$ space, we may define its circulation over any closed path $C$ as

\begin{equation}\label{kc}
\kappa_{C}=\oint_{C}\boldsymbol{\ell}\cdot d\boldsymbol{k}_{\|}.
\end{equation}
$\kappa_{C}$ will be quantized when $\boldsymbol{\mathcal A}_r-\boldsymbol{\mathcal A}_i=0$.
As discussed in Ref.~\cite{Yliu2020}, this condition can be satisfied when the incident medium is a trivial metal, where the Berry connection $\boldsymbol{\mathcal A}_{i(r)}$ vanishes for the band at Fermi level.
It also holds if the incident medium has mirror symmetry $M_x$, making $\boldsymbol{\mathcal A}_r$ and $\boldsymbol{\mathcal A}_i$ cancel out each other.
For such cases, the CAS is quantized
\begin{equation}\label{QCL}
\kappa_{C}/(2\pi)=-\frac{1}{2\pi}\oint_{C}\frac{\partial \phi}{\partial\boldsymbol{k}_{\|}}\cdot d\boldsymbol{k}_{\|}\ \in\mathbb{Z}.
\end{equation}
From the above expression, one can see that this quantized number $\kappa_{C}/(2\pi)$ is just the (negative) winding number of the phase angle of reflection amplitude when we go around the loop $C$.

An interesting scenario is to have the incident medium trivial, e.g., a simple metal or just vacuum, then the anomalous shift and its quantized CAS would help to unveil information of the target medium. Particularly, it was shown in Ref.~\cite{Yliu2020} that
if the target medium contains a Weyl point with charge $N$, then it would produce
\begin{equation}\label{kappaW}
  \kappa_{C}=-2\pi N,
\end{equation}
when $C$ encloses the projection of the Weyl point in the interface momentum plane.
This offers a method to probe the Weyl point and its topological charge.

\begin{figure}[t]
	\includegraphics[width=8.7cm]{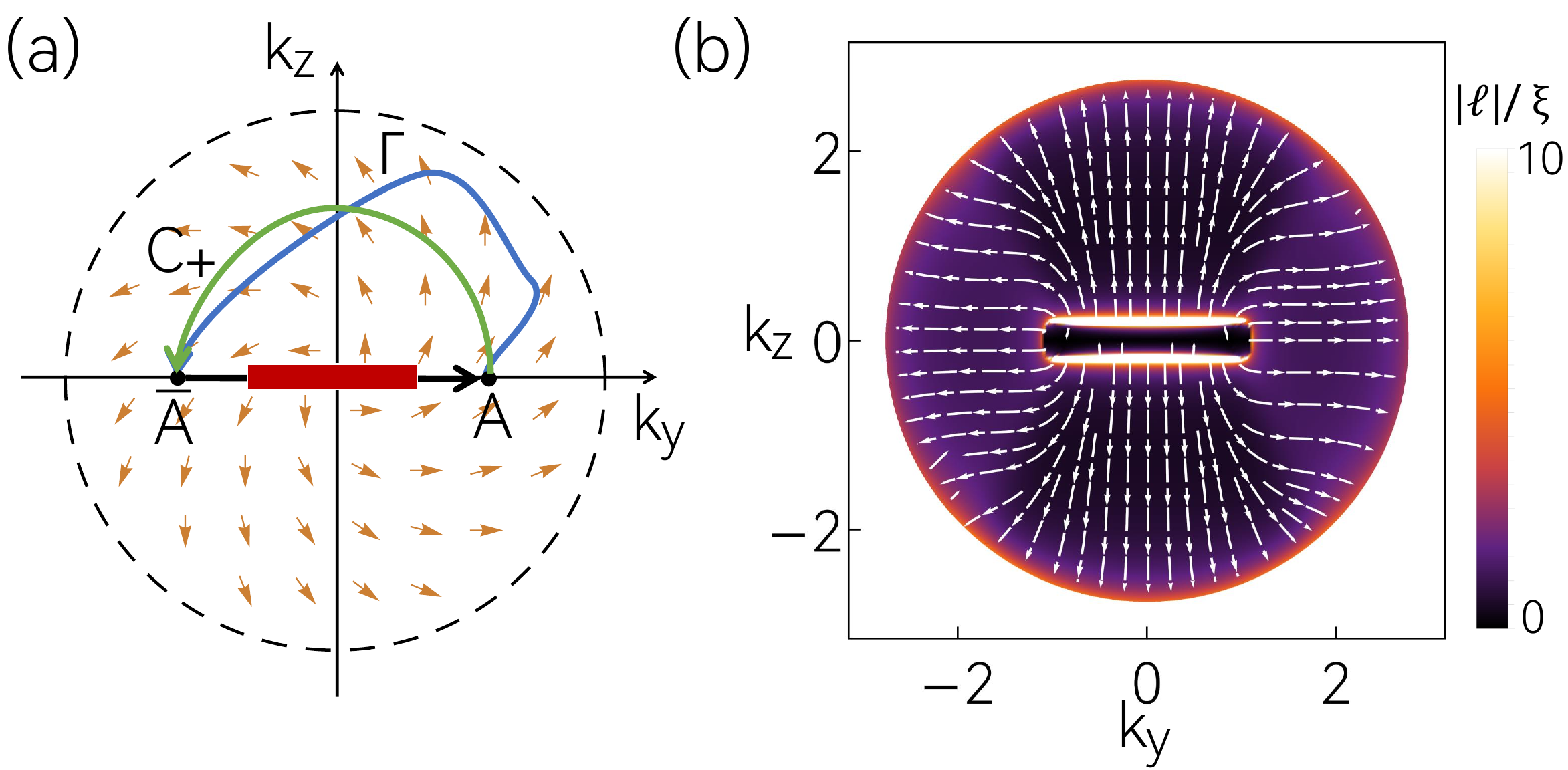}
	\caption{(a) Illustration of CAS $\kappa_s$ over a semicircle $C_+$ with symmetric ends $A$ and $\bar A$. The red line makes the range of the nodal ring. $\Gamma$ is another path with arbitrary shape but with the two ends $A$ and $\bar A$ fixed.  (b) Numerical results for the shift vector field $\boldsymbol{\ell}$ in the interface momentum plane for a ring with a trivial charge. The magnitude and the
		direction of the shift vectors are indicated by the color map and the arrows, respectively. Here, we take $n=0$, $E_F=0.2$, $U=8$, $V=16$, $A=1$, $B=1$, $C=-1$ and $D=-1$. $\xi$ is a unit of length, defined as $\xi=A/B$.
		\label{semiCAS}}
\end{figure}

\subsection{CAS on a semicircle}

Equation (\ref{kappaW}) was found for Weyl points. We will see below that it also applies to the mirror-protected nodal rings. However, as we mentioned in Sec.~\ref{secII}. that these rings have $N=0$, the relation (\ref{kappaW}) does not give much information. Evidently, it is desirable to capture the refine charge $\chi_h$ in certain features of the shift vector field, hopefully with a relation similar to (\ref{kappaW}).

We find this is indeed possible, if some additional symmetry is imposed. Recall that $\chi_h$ is defined on a hemisphere.
In the interface momentum plane, it is then natural to consider a loop that involves the semicircle, as shown in Fig.~\ref{semiCAS}(a).
Here, following the setup in Sec.~\ref{secII}, we consider a nodal ring protected by $M_z$ symmetry and lying in the $k_z=0$ plane.
The loop $C$ considered here consists of a semicircle $C_+$ and a straight path on the $k_z=0$ line. The two ends of the straight path are labeled as $A$ and $\bar{A}$, with opposite $k_y$ values. To probe the nodal ring, one should let $\bar{A}A$ contain the spread of the ring. Then, we have
\begin{equation}
\kappa_{C}=\left(\int_{C_+}+\int_{\bar{A}A}\right) \boldsymbol{\ell}\cdot d\boldsymbol{k}_{\|}.
\end{equation}
Like in the definition of $\chi_h$, we want to have the contribution from the $\bar{A}A$ segment vanish.
This requires an additional symmetry for the system, e.g., either $C_{2x}$ or $M_y$ symmetry will suffice.
The nodal ring model (\ref{HH}) has both symmetries. Under such a condition, we would have
\begin{equation}\label{ks}
  \kappa_s\equiv\int_{C_+}\boldsymbol{\ell}\cdot d\boldsymbol{k}_{\|}=\kappa_C
\end{equation}
also take a quantized value.

Since $\kappa_s$ is quantized, its value should not change under small perturbation to the path $s$, as long as does not cross topologically charged degeneracies such as Weyl points. For example, in  Fig.~\ref{semiCAS}(a), $\kappa_s$ computed for the path $\Gamma$ should equal to that on $C_+$, provided that there is no degeneracy in the shaded region.

Now, we have managed to adapt the original definition of CAS so it may apply to mirror-protected nodal rings.
The refined quantity $\kappa_s$ gives a quantized value. The question is: Can this value be nonzero? And how does it related to the topological charge $\chi_h$ of vortex rings? We shall answer these questions in the following.

\section{Shift for nodal rings}
Our setup has been illustrated in Fig.~\ref{junction}. The incident medium at $x<0$ is taken to be a trivial, described by the model
 \begin{equation}
 	H_1=\Big(\frac{k^2}{2m}+U\Big)\sigma_z-V,
 \end{equation}
where $m$, $U$, and $V$ are real model parameters, and we take their values to be positive. $U$ denotes a gap between the two bands, and $V$ is a relative potential energy shift across the interface.
The target medium at $x>0$ is taken to be the nodal-ring semimetal, with its model $H_2$ described by Eq.~(\ref{HH}).

The incident mode has the magnitude of its interface momentum limited by
\begin{equation}
  k_F=\sqrt{2m(E_F-U+V)},
\end{equation}
where $E_F$ is the Fermi energy. To have the nodal ring close to Fermi level, we should have $E_F$ being small ($E_F=0$ corresponds to the case with the nodal ring exactly at Fermi level).  This means the shift vector field is also limited within a circular region of this radius
in the interface momentum plane. For a given interface momentum in this region, i.e., $k_\|<k_F$, the corresponding incident mode is given by
\begin{equation}
  \psi_{i}=\left(\begin{array}{c}
		1\\
		0
	\end{array}\right)e^{ik_{i}x}e^{ik_y y+ik_z z},
\end{equation}
and there is only one propagating reflected mode
\begin{equation}
  \psi_{r}^{1}=\left(\begin{array}{c}
		1\\
		0
	\end{array}\right)e^{-ik_{i}x}e^{ik_y y+ik_z z},
\end{equation}
where
\begin{equation}
  k_i=\sqrt{2m(E_F-U+V)- k_{\|}^2}.
\end{equation}
In the following, we shall solve the anomalous shift for nodal rings with different topological charges.

\subsection{Conventional ring}\label{SCR}
Let's first consider the case of the conventional ring with zero topological charge $\chi_h$.
This corresponds to the model (\ref{model1}) with $n=0$. To be specific, we take $A, B>0$ and $C,D<0$, so the two bands cross at a single nodal ring.
For this case, the transmitted modes in $x>0$ region can be found as
\begin{equation}\label{st}
	\psi_{t}^{\pm}=\frac{1}{\sqrt{2}}\left(\begin{array}{c}
		1\\
		e^{\pm i\varphi}
	\end{array}\right)e^{ik_{t}^{\pm}x},
\end{equation}
where
\begin{equation}
  k_{t}^{\pm}=\mp\sqrt{-\frac{D}{B}-k_y^2-\frac{C}{B} k_z^2\mp 2 \sqrt{E_F^2-A^2k_z^2}},
\end{equation} and
\begin{equation}
  \varphi=\operatorname{sgn}(k_z)\arccos(\frac{E_F}{Ak_z}).
\end{equation}

The scattering amplitudes are solved from the continuity of the wave function at the interface $x=0$. In particular, the reflection amplitude can be found as
\begin{equation}\label{r1}
	r=\frac{e^{2 i \varphi}(k_i-2mB k_t^{-})-k_i+2  mB k_t^{+}}{e^{2 i \varphi}(k_i+2mB k_t^{-})-k_i-2 mB k_t^{+}}.
\end{equation}

It is difficult to directly compute $\kappa_s$ from the integral (\ref{ks}). However, noticing that the model here has no other band degeneracies, the value of $\kappa_s$ should not change when we vary the two ends of the semicircle $C_+$, as long as they enclose the projection of the ring. Such topological invariance allows us to evaluate $\kappa_s$ using some simple limiting case. For example, we may take $B=C$ in the model (\ref{model1}) and take $E_F+V\gg U$.
The latter makes $k_F$ much larger than the radius $R$ of the ring, and so we may consider a semicircle $C_+$ with radius $\rho\gg R$. For such a limiting case, one finds that $k_t^{\pm}\approx i \rho$ is purely imaginary, and the reflection amplitude
\begin{equation}\label{r1}
	r\approx\frac{k_i-i 2m B\rho}{k_i+i 2m B\rho}.
\end{equation}
It can be observed that $r$ and $\phi=\arg (r)$ take constant values across $C_+$. Therefore, we have
\begin{equation}
\kappa_s=0,
\end{equation}
when the ring has trivial topological charge $\chi_h$.

In Fig.~\ref{semiCAS}(b), we plot the numerically calculated shift vector field in the interface momentum plane. The color map reflects the
magnitude of the shift vector.
One clearly observes that the anomalous shift is greatly enhanced near the (projection of) nodal ring.
From the pattern in Fig.~\ref{semiCAS}(b), one can see the CAS $\kappa_s$ vanishes. In addition, there is a notable feature that the shift is almost perpendicular to the ring's plane, i.e., in the $k_z$ direction, on the two sides of the ring. To understand this, note that for $\bm k_\|$ in the vicinity of the ring,
$k_{t}^{\pm}$ is real. The reflection amplitude
\begin{equation}\label{r3}
	r\approx -\frac{2m B k_t^{+}+ik_i \tan \varphi}{2m B k_t^{+}-ik_i \tan \varphi}
\end{equation}
has a unit modulus. In such a case, we may write
\begin{equation}\label{l2}
	\boldsymbol{\ell}=-\frac{\partial \phi}{\partial \boldsymbol{k}_\|}\approx-i\frac{1}{r}\frac{\partial r}{\partial \boldsymbol{k}_\|}.
\end{equation}
For $E_F\approx 0$, one finds that
\begin{equation}
  \ell_z/\ell_y\approx \partial_{k_z} r/\partial_{k_y} r\sim \Lambda \frac{R}{k_z},
\end{equation}
where $\Lambda$ is some number of  order unity. Therefore, near the nodal ring and for small $k_z$, the $z$ component of the shift $\ell_z$ dominates over the $y$ component, and the shift is almost perpendicular to the ring.
\begin{figure}[t]
	\includegraphics[width=8.7cm]{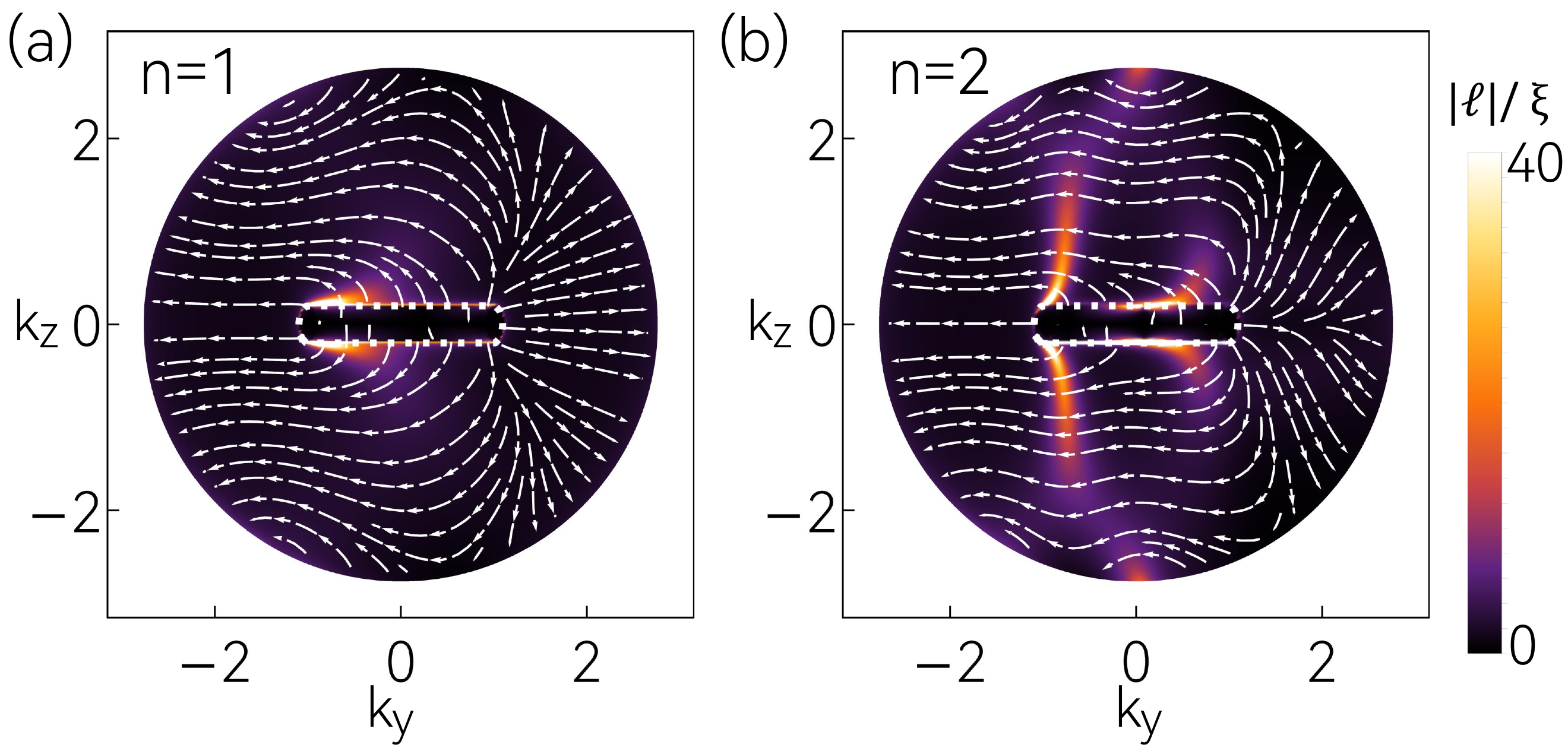}
	\caption{Calculated shift vector fields for vortex rings with (a) $n=1$ and (b) $n=2$. The magnitude and the
		direction of the shift vectors are indicated by the color map and the arrows, respectively. The white dashed lines indicate the range of projected nodal rings.
		Here, we set $E_F=0.2$, $U=8$, $V=16$, $A=1$, $B=1$, $C=-1/2$, and $D=-1$.
		\label{VNLvector2}}
\end{figure}

\subsection{Vortex  ring}
Next, we consider the case of a vortex nodal ring. In model (\ref{HH}), we take $n\geq 1$, $A, B>0$ and $C, D<0$. As discussed in Sec.~\ref{secII}, such a ring has  a nontrivial topological charge $\chi_h=-n$. We will see that this charge manifests in a nonzero CAS $\kappa_s$.

It is difficult to obtain analytic expression of the anomalous shift for this case, hence we perform numerical investigations. In Figs.~\ref{VNLvector2}~(a) and (b), we show the calculated shift vector field for $n=1$ and $n=2$ cases, respectively.
Again, one observes that the shift is generally enhanced at the projection of the ring. The field pattern is symmetry about the horizontal line $k_z=0$, as required by the $M_z$ mirror symmetry.

\begin{figure}[t]
	\includegraphics[width=8.7cm]{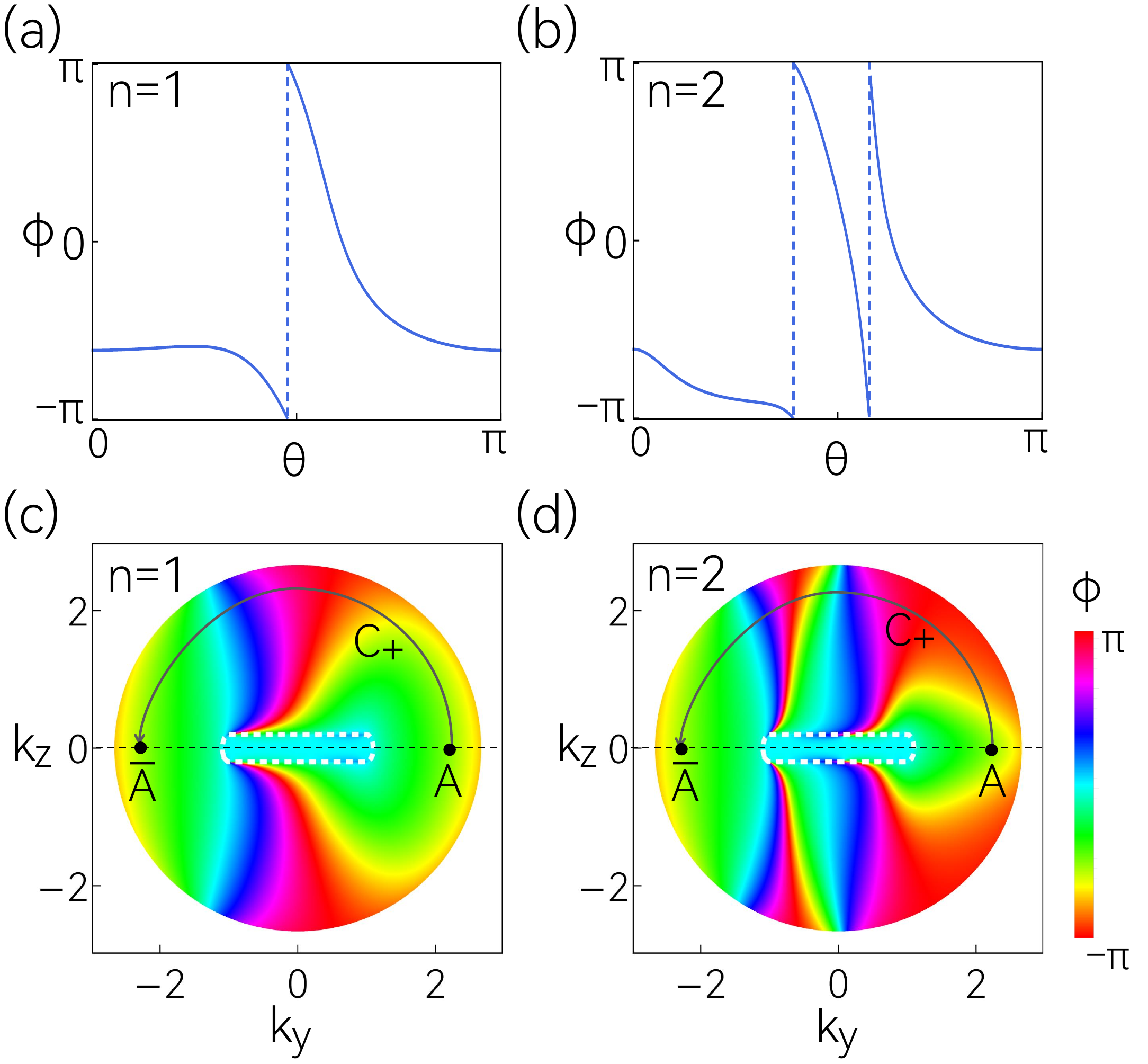}
	\caption{(a,b) Evolution of phase angle $\phi=\arg(r)$ on the semicircle $C_+$ (from $\theta=0$ to $\theta=\pi$). (c,d) Distribution of $\arg(r)$ in the interface momentum plane. (a,c) are for nodal ring with $n=1$, and (b,d) are for $n=2$ case. Here, we set $E_F=0.2$, $U=8$, $V=16$, $A=1$, $B=1$, $C=-1/2$, and $D=-1$.
		\label{argfig}}
\end{figure}

The CAS $\kappa_s$ cannot be easily seen from Figs.~\ref{VNLvector2}. Note that if we take a semicircular path $C_+$ labeled by polar angle $\theta$ (with the two ends of $C_+$ enclosing the ring), then $\kappa_s$ would corresponding to the (negative of) winding number of phase angle $\phi = \arg r(\theta)$ over $2\pi$, when $\theta$ increases from $0$ to $\pi$. This is similar to the method for the evaluation of $\chi_h$ in Fig.~\ref{nodalstructure}. We plot $\phi$ as a function of $\theta$ for the semicircles shown in Figs.~\ref{argfig}(a,b). One finds that $\kappa_s$ is proportional to $n$ and is given by $\kappa_s=2\pi n$. Since $\chi_h=-n$, this establishes the relation $\kappa_s=-2\pi \chi_h$ in Eq.~(\ref{rel}).
We have also numerically studied the cases with higher values of $n$, and the result confirms the general validity of this relation.

Another perhaps more intuitive approach is to visualize $\kappa_s$ is to plot the phase angle distribution $\phi$ in the interface momentum plane. It is a scalar field on this plane, valued in $[-\pi,\pi)$ with the two ends identified.
In Figs.~\ref{argfig}(c,d), we show the plots for $n=1$ and $n=2$ cases, using a circular color-map for the value of $\phi$. One can clearly see that when going around a semicircle (or any path) from $A$ to $\bar A$ point, the angle $\phi$ will wind around $2\pi$ (i.e., traverse all the colors) $n$ times, consistent with the discussion above.

\begin{figure*}[t]
	\includegraphics[width=16cm]{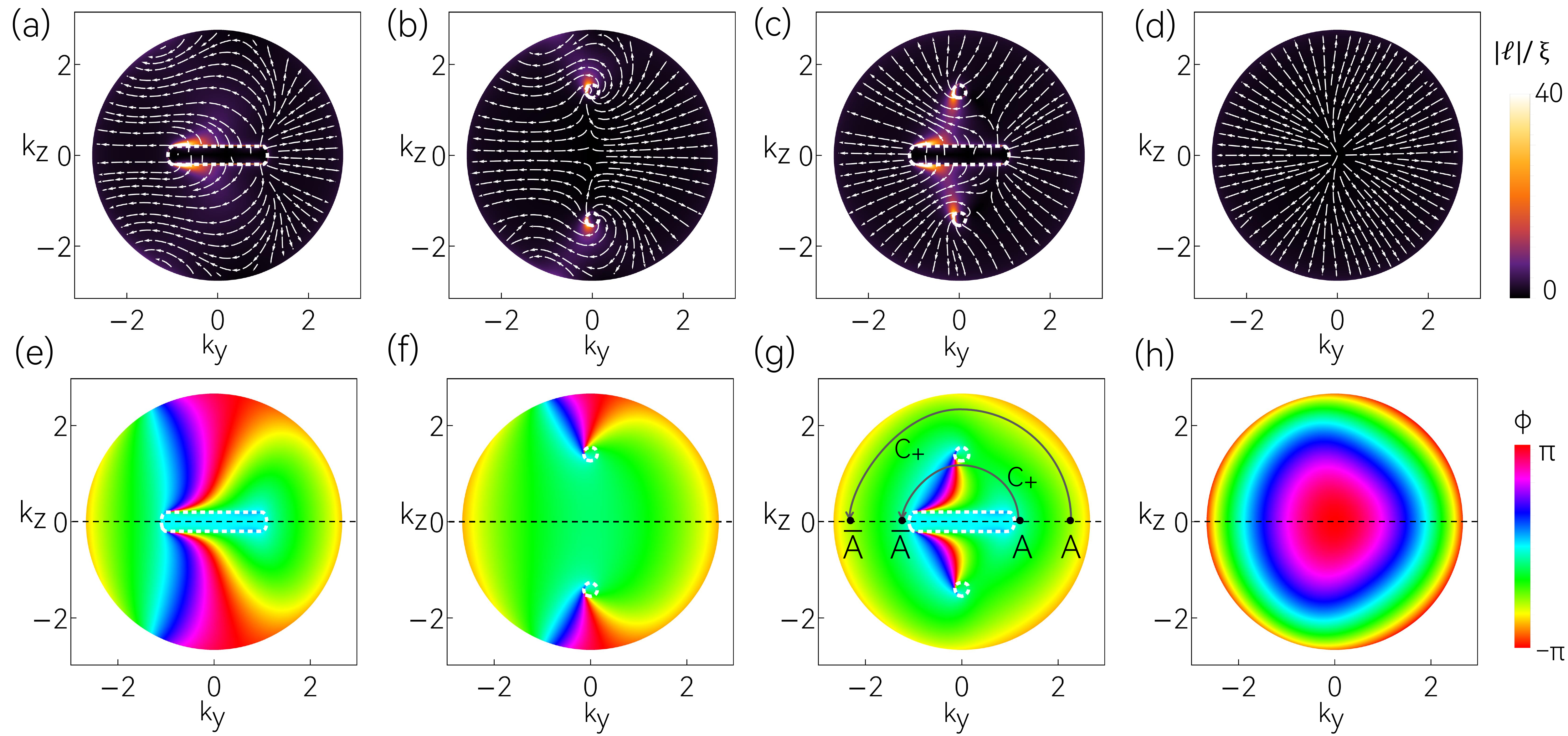}
	\caption{
Calculated shift vector fields and  distribution of $\arg(r)$ for $n=1$ in the four phases of phase diagram Fig.~\ref{nodalstructure}(d). The topological features are marked with white dashed lines. (a,e) are for the nodal ring phase. (b,f) are for the Weyl point phase. (c,g) are for the nodal ring plus Weyl point phase. (d,h) are for the gapped phase. In the calculation, $E_F=0.2$, $U=8$, $V=16$, $A=1$, and $B=1$.  In addition, we set $C=-1/2$ and $D=-1$ in (a,e); $C=-1/2$ and $D=1$ in (b,f); $C=1/2$ and $D=-1$ in (c,g); $C=1/2$ and $D=1$ in (d,h).
		\label{vF4}}
\end{figure*}

Finally, we may also obtain the value of $\kappa_s$ by an analysis of the limiting case. For example, take the $n=1$ vortex ring.  Similar to the argument in Sec.~\ref{SCR}, we consider the limiting case of a semicircle having a large radius $\rho$. In this case, the transmitted states in the target medium can be cast into the form

\begin{equation}
	\psi_{t}^{\pm}=\frac{1}{\sqrt{\eta_{\pm}^{2}+1}}\left(\begin{array}{c}
		1\\
		i\eta_{\pm}
	\end{array}\right)e^{k_{t}^{\pm}x},
\end{equation}
where $k_{t}^{\pm}=\sqrt{k_{y}^{2}+\tau_{\pm}^{2}k_{z}^{2}}$,
$
  \eta_{\pm}\approx\pm\frac{k_{t}^{\pm}+k_{y}}{\tau_{\pm}k_z}
$,
$\tau_{+}\approx C/A$, $\tau_- \approx A/B$, and we have assumed $|C|$ is much smaller than $A$ and $B$. The reflection amplitude has a unit modulus, and hence can be written in the form of
\begin{equation}\label{Ra}
	r=-\frac{\alpha-i\beta}{\alpha+i\beta},
\end{equation}
where $\alpha=2Bm(k_{t}^{+}\eta_{-}-k_{t}^{-}\eta_{+})$, and $\beta={k_{i}}(\eta_{-}-\eta_{+})$ are real functions of  $\theta$. If $\phi=\text{arg}(r)$ has a nonzero winding number as $\theta$ varies from $0$ to $\pi$, it must cross any given value between $-\pi$ and $\pi$. For example, taking $\pi$ as a reference value, we may check whether $\phi=\pi$ has any solution $\theta\in[0,\pi]$. From the above result, to have $\phi=\pi$, we must have $\beta(\theta)=0$. And this leads to the following equation for $\theta$:
\begin{equation}\label{crossingnum}\begin{split}
A^2\cos\theta &+ C \sqrt{B^2\cos^2\theta+ A^2 \sin^2 \theta} \\
&\qquad + A \sqrt{A^2 \cos^2 \theta + C^2 \sin^2 \theta}=0.
\end{split}
\end{equation}
One can show that this equation has a unique solution $\theta=\pi/2$ for $\theta\in[0,\pi]$. Now, since $\phi(0)=\phi(\pi)$ and $\phi(\theta)$ crosses $\pi$ only once, the winding number of $\phi(\theta)$ has to be unity. This offers an alternative way to confirm our result obtained above.

\section{Topological phase transition}

So far, our investigation has been focused on the nodal-ring phase, i.e., the lower left corner of the phase diagram Fig.~\ref{nodalstructure}(d).
It can be directly extended to other phases in Fig.~\ref{vF4}.
To be specific, consider the $n=1$ case. The cases with larger $n$ values are similar.

As a first example, starting from the nodal-ring phase with $C<0$ and $D<0$ shown in Fig.~\ref{vF4}(a), we increase $D$ and make a topological phase transition into the Weyl semimetal phase with $C<0$ and $D>0$. In this process, the nodal ring shrinks to a point and
then transforms into a pair of Weyl points. The corresponding shift vector field is shown in Fig.~\ref{vF4}(b)
One can see two hot spots with enhanced anomalous shift at the (projected) position of the two Weyl points.
Fig.~\ref{vF4}(f) shows the scalar field $\phi$ in the interface momentum plane. Evidently, for any semicircle that covers the Weyl point, the CAS $\kappa_s$ would still be $-2\pi$, same as the nodal ring case in Fig.~\ref{vF4}(e) (And for any loop enclosing the lower Weyl point, the CAS should be $2\pi$.)

Next, consider the transition from the nodal-ring phase (with $C<0$ and $D<0$) to the hybrid phase (with $C>0$ and $D<0$) by increasing $C$.
As discussed, two Weyl points emerges at infinity right after $C$ crosses zero. At this moment, the CAS on a semicircle $C_+$ with finite radius still has a value of $-2\pi$, as shown in Fig.~\ref{vF4}(c), since the vortex nodal ring with $\chi_h=-1$ is maintained. With further increase of $C$, the two Weyl points move closer to the origin. Once a Weyl point crosses $C_+$, the CAS $\kappa_s$ on $C_+$ jumps to zero, as shown in Fig.~\ref{vF4}(g). This can be understood from the cancelation of the charges of the vortex ring and the Weyl point (the Weyl point has $\chi_h=1$), and the relation (\ref{rel}) still holds.

Finally, for completeness, in Figs.~\ref{vF4}(d,h), we also present the results for the gapped phase with $C, D>0$. One can see that $\kappa_s$ is always zero for this trivial phase.

\begin{figure}[t]
	\includegraphics[width=9cm]{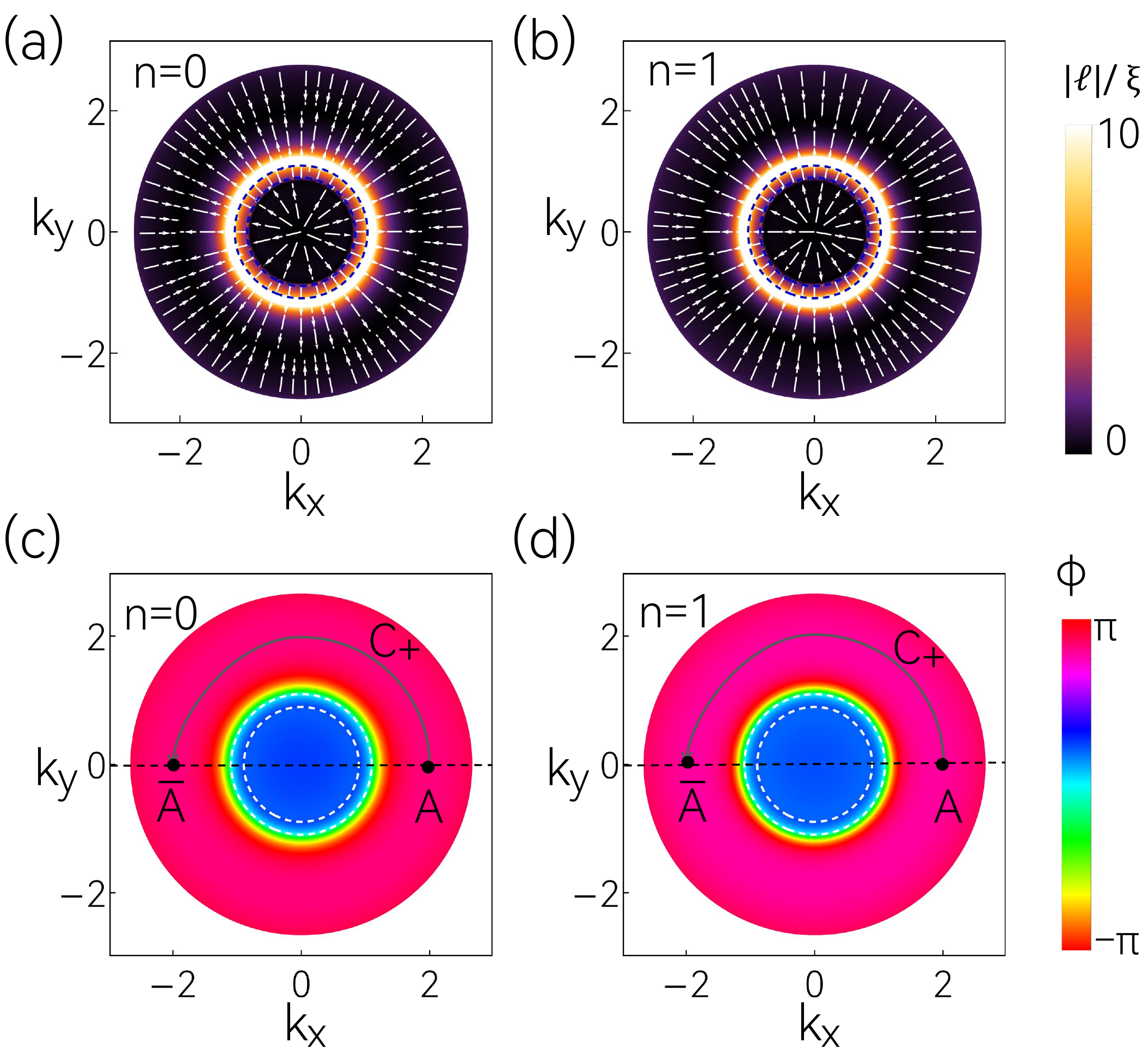}
	\caption{(a,b) Calculated shift vector fields and (c,d) distribution of $\arg(r)$ for the configuration with nodal rings in the $k_z=0$ plane, parallel to the interface. The dashed curves indicate the projection of the nodal rings. Here, we set $E_F=0.2$, $U=8$, $V=16$, $A=1$, $B=1$, $C=-1$, and $D=-1$.
		\label{VNLvectorz}}
\end{figure}

\section{Discussion and Conclusion}

We have unveiled a physical manifestation of the topological charge $\chi_h$ of mirror-protected nodal rings. In the considered setup, the mirror plane that contains the nodal ring is normal to the surface of the nodal-ring medium.
This is certainly the natural choice, since the topological charge $\chi_h$ is defined on a hemisphere based on the mirror plane. If the anomalous shift is probed on another surface that is parallel to the mirror plane, then generally one cannot expect to have a CAS manifesting the charge $\chi_h$. This point is confirmed in Fig.~\ref{VNLvectorz}, which shows the result for such a configuration. It is shown that the value of $\chi_h$ does not produce any qualitative change here.  Nevertheless, one can see that the shift is still greatly enhanced at the projection of the nodal ring. Therefore, detecting the shift still offers a way to probe the shape of the nodal ring in momentum space.

In conclusion, we have studied the anomalous shift in scattering from mirror-protected nodal-ring semimetals. We find that the nodal ring generally enhances the shift magnitude, so mapping out the shift vector field can help to probe the extent and the shape of the ring. More importantly, we unveil a manifestation of the topological charge $\chi_h$ for such rings. By adapting the definition of CAS for mirror symmetric systems, we find the quantized value of $\kappa_s$ encodes the information of $\chi_h$ in a very nice relationship.

This work offers a new approach for analyzing the topological properties of nodal rings and demonstrates the rich information contained in anomalous scattering shifts. Beyond Weyl points and nodal rings, there are many other possible topological band nodal structures, as classified in recent works~\cite{yu2022,LiuGB2022,Zhang2022,Zhang2023}. It is natural to anticipate that there would be fascinating consequences in the anomalous scattering shift arising from topological semimetals containing those structures.
This will be an interesting direction for future research.

\begin{acknowledgements}
The authors thank D. L. Deng for valuable discussions.
This work is supported by the National Natural Science Foundation of China (Grants No. 12474040
and No. 12374039), the Hebei Natural Science Foundation A2023202009, and the HK PolyU start-up grant (P0057929).
\end{acknowledgements}

\bibliographystyle{apsrev4-2}
\bibliography{cite}

@article{Bliokh_2013,
doi = {10.1088/2040-8978/15/1/014001},
url = {https://dx.doi.org/10.1088/2040-8978/15/1/014001},
year = {2013},
month = {jan},
publisher = {IOP Publishing},
volume = {15},
number = {1},
pages = {014001},
author = {K Y Bliokh and A Aiello},
title = {Goos–Hänchen and Imbert–Fedorov beam shifts: an overview},
journal = {Journal of Optics}
}

@ARTICLE{Goos1947,
       author = {{Goos}, F. and {H{\"a}nchen}, H.},
        title = "{Ein neuer und fundamentaler Versuch zur Totalreflexion}",
      journal = {Annalen der Physik},
         year = 1947,
        month = jan,
       volume = {436},
       number = {7},
        pages = {333-346},
          doi = {10.1002/andp.19474360704},
       adsurl = {https://ui.adsabs.harvard.edu/abs/1947AnP...436..333G}
}

@article{IF1955,
  title = {IF1},
  author = {F. I. Fedorov, Dokl. Akad.},
  journal = {Nauk SSSR},
  volume = {105},
  pages = {465},
  year = {1955},
}

@article{IF1972,
  title = {Calculation and Experimental Proof of the Transverse Shift Induced by Total Internal Reflection of a Circularly Polarized Light Beam},
  author = {Imbert, Christian},
  journal = {Phys. Rev. D},
  volume = {5},
  issue = {4},
  pages = {787--796},
  numpages = {0},
  year = {1972},
  month = {Feb},
  publisher = {American Physical Society},
  doi = {10.1103/PhysRevD.5.787},
  url = {https://link.aps.org/doi/10.1103/PhysRevD.5.787}
}

@article{Miller1972,
  title = {Shifts of Electron Beam Position Due to Total Reflection at a Barrier},
  author = {Miller, Stanley C. and Ashby, Neil},
  journal = {Phys. Rev. Lett.},
  volume = {29},
  issue = {11},
  pages = {740--743},
  numpages = {0},
  year = {1972},
  month = {Sep},
  publisher = {American Physical Society},
  doi = {10.1103/PhysRevLett.29.740},
  url = {https://link.aps.org/doi/10.1103/PhysRevLett.29.740}
}

@article{Beenakker2009,
  title = {Quantum Goos-H\"anchen Effect in Graphene},
  author = {Beenakker, C. W. J. and Sepkhanov, R. A. and Akhmerov, A. R. and Tworzyd\l{}o, J.},
  journal = {Phys. Rev. Lett.},
  volume = {102},
  issue = {14},
  pages = {146804},
  numpages = {4},
  year = {2009},
  month = {Apr},
  publisher = {American Physical Society},
  doi = {10.1103/PhysRevLett.102.146804},
  url = {https://link.aps.org/doi/10.1103/PhysRevLett.102.146804}
}

@article{Yu2019,
  title = {Anomalous spatial shifts in interface electronic scattering},
  author = {Yu, Zhi-Ming and Liu, Ying and Yang, Shengyuan A.},
  journal = {Frontiers of Physics},
  volume = {14},
  issue = { 3},
  year = {2019},
  month = {Apr},
  pages={33402},
  doi = {10.1007/s11467-019-0882-7},
  url = {https://doi.org/10.1007/s11467-019-0882-7}
}

@article{Chiu2016,
  title = {Classification of topological quantum matter with symmetries},
  author = {Chiu, Ching-Kai and Teo, Jeffrey C. Y. and Schnyder, Andreas P. and Ryu, Shinsei},
  journal = {Rev. Mod. Phys.},
  volume = {88},
  issue = {3},
  pages = {035005},
  numpages = {63},
  year = {2016},
  month = {Aug},
  publisher = {American Physical Society},
  doi = {10.1103/RevModPhys.88.035005},
  url = {https://link.aps.org/doi/10.1103/RevModPhys.88.035005}
}

@article{Armitage2018,
  title = {Weyl and Dirac semimetals in three-dimensional solids},
  author = {Armitage, N. P. and Mele, E. J. and Vishwanath, Ashvin},
  journal = {Rev. Mod. Phys.},
  volume = {90},
  issue = {1},
  pages = {015001},
  numpages = {57},
  year = {2018},
  month = {Jan},
  publisher = {American Physical Society},
  doi = {10.1103/RevModPhys.90.015001},
  url = {https://link.aps.org/doi/10.1103/RevModPhys.90.015001}
}

@article{Wan2011,
  title = {Topological semimetal and Fermi-arc surface states in the electronic structure of pyrochlore iridates},
  author = {Wan, Xiangang and Turner, Ari M. and Vishwanath, Ashvin and Savrasov, Sergey Y.},
  journal = {Phys. Rev. B},
  volume = {83},
  issue = {20},
  pages = {205101},
  numpages = {9},
  year = {2011},
  month = {May},
  publisher = {American Physical Society},
  doi = {10.1103/PhysRevB.83.205101},
  url = {https://link.aps.org/doi/10.1103/PhysRevB.83.205101}
}

@article{ZW2011,
  title = {Valley-Dependent Brewster Angles and Goos-H\"anchen Effect in Strained Graphene},
  author = {Wu, Zhenhua and Zhai, F. and Peeters, F. M. and Xu, H. Q. and Chang, Kai},
  journal = {Phys. Rev. Lett.},
  volume = {106},
  issue = {17},
  pages = {176802},
  numpages = {4},
  year = {2011},
  month = {Apr},
  publisher = {American Physical Society},
  doi = {10.1103/PhysRevLett.106.176802},
  url = {https://link.aps.org/doi/10.1103/PhysRevLett.106.176802}
}

@article{Jiang2015,
  title = {Topological Imbert-Fedorov Shift in Weyl Semimetals},
  author = {Jiang, Qing-Dong and Jiang, Hua and Liu, Haiwen and Sun, Qing-Feng and Xie, X. C.},
  journal = {Phys. Rev. Lett.},
  volume = {115},
  issue = {15},
  pages = {156602},
  numpages = {5},
  year = {2015},
  month = {Oct},
  publisher = {American Physical Society},
  doi = {10.1103/PhysRevLett.115.156602},
  url = {https://link.aps.org/doi/10.1103/PhysRevLett.115.156602}
}

@article{Yang2015,
  title = {Chirality-Dependent Hall Effect in Weyl Semimetals},
  author = {Yang, Shengyuan A. and Pan, Hui and Zhang, Fan},
  journal = {Phys. Rev. Lett.},
  volume = {115},
  issue = {15},
  pages = {156603},
  numpages = {5},
  year = {2015},
  month = {Oct},
  publisher = {American Physical Society},
  doi = {10.1103/PhysRevLett.115.156603},
  url = {https://link.aps.org/doi/10.1103/PhysRevLett.115.156603}
}

@article{Yliu2017,
  title = {Transverse shift in Andreev reflection},
  author = {Liu, Ying and Yu, Zhi-Ming and Yang, Shengyuan A.},
  journal = {Phys. Rev. B},
  volume = {96},
  issue = {12},
  pages = {121101},
  numpages = {5},
  year = {2017},
  month = {Sep},
  publisher = {American Physical Society},
  doi = {10.1103/PhysRevB.96.121101},
  url = {https://link.aps.org/doi/10.1103/PhysRevB.96.121101}
}

@article{1Yliu2018,
  title = {Goos-H\"anchen-like shifts at a metal/superconductor interface},
  author = {Liu, Ying and Yu, Zhi-Ming and Jiang, Hua and Yang, Shengyuan A.},
  journal = {Phys. Rev. B},
  volume = {98},
  issue = {7},
  pages = {075151},
  numpages = {8},
  year = {2018},
  month = {Aug},
  publisher = {American Physical Society},
  doi = {10.1103/PhysRevB.98.075151},
  url = {https://link.aps.org/doi/10.1103/PhysRevB.98.075151}
}

@article{Yu2018,
  title = {Unconventional Pairing Induced Anomalous Transverse Shift in Andreev Reflection},
  author = {Yu, Zhi-Ming and Liu, Ying and Yao, Yugui and Yang, Shengyuan A.},
  journal = {Phys. Rev. Lett.},
  volume = {121},
  issue = {17},
  pages = {176602},
  numpages = {5},
  year = {2018},
  month = {Oct},
  publisher = {American Physical Society},
  doi = {10.1103/PhysRevLett.121.176602},
  url = {https://link.aps.org/doi/10.1103/PhysRevLett.121.176602}
}

@article{2Yliu2018,
  title = {Transverse shift in crossed Andreev reflection},
  author = {Liu, Ying and Yu, Zhi-Ming and Liu, Jie and Jiang, Hua and Yang, Shengyuan A.},
  journal = {Phys. Rev. B},
  volume = {98},
  issue = {19},
  pages = {195141},
  numpages = {10},
  year = {2018},
  month = {Nov},
  publisher = {American Physical Society},
  doi = {10.1103/PhysRevB.98.195141},
  url = {https://link.aps.org/doi/10.1103/PhysRevB.98.195141}
}

@article{Yliu2020,
  title = {Quantized Circulation of Anomalous Shift in Interface Reflection},
  author = {Liu, Ying and Yu, Zhi-Ming and Xiao, Cong and Yang, Shengyuan A.},
  journal = {Phys. Rev. Lett.},
  volume = {125},
  issue = {7},
  pages = {076801},
  numpages = {6},
  year = {2020},
  month = {Aug},
  publisher = {American Physical Society},
  doi = {10.1103/PhysRevLett.125.076801},
  url = {https://link.aps.org/doi/10.1103/PhysRevLett.125.076801}
}

@article{WU2018,
  title = {Nodal surface semimetals: Theory and material realization},
  author = {Wu, Weikang and Liu, Ying and Li, Si and Zhong, Chengyong and Yu, Zhi-Ming and Sheng, Xian-Lei and Zhao, Y. X. and Yang, Shengyuan A.},
  journal = {Phys. Rev. B},
  volume = {97},
  issue = {11},
  pages = {115125},
  numpages = {11},
  year = {2018},
  month = {Mar},
  publisher = {American Physical Society},
  doi = {10.1103/PhysRevB.97.115125},
  url = {https://link.aps.org/doi/10.1103/PhysRevB.97.115125}
}

@article{YU2022,
title = {Encyclopedia of emergent particles in three-dimensional crystals},
journal = {Science Bulletin},
volume = {67},
number = {4},
pages = {375-380},
year = {2022},
issn = {2095-9273},
doi = {https://doi.org/10.1016/j.scib.2021.10.023},
url = {https://www.sciencedirect.com/science/article/pii/S2095927321006927},
author = {Zhi-Ming Yu and Zeying Zhang and Gui-Bin Liu and Weikang Wu and Xiao-Ping Li and Run-Wu Zhang and Shengyuan A. Yang and Yugui Yao}
}

@article{Do2022,
  title = {Effects of surface potentials on Goos-H\"anchen and Imbert-Fedorov shifts in Weyl semimetals},
  author = {Dongre, Ninad Kailas and Roychowdhury, Krishanu},
  journal = {Phys. Rev. B},
  volume = {106},
  issue = {7},
  pages = {075414},
  numpages = {11},
  year = {2022},
  month = {Aug},
  publisher = {American Physical Society},
  doi = {10.1103/PhysRevB.106.075414},
  url = {https://link.aps.org/doi/10.1103/PhysRevB.106.075414}
}

@article{shi2019,
  title = {Shift vector as the geometric origin of beam shifts},
  author = {Shi, Li-kun and Song, Justin C. W.},
  journal = {Phys. Rev. B},
  volume = {100},
  issue = {20},
  pages = {201405},
  numpages = {5},
  year = {2019},
  month = {Nov},
  publisher = {American Physical Society},
  doi = {10.1103/PhysRevB.100.201405},
  url = {https://link.aps.org/doi/10.1103/PhysRevB.100.201405}
}

@article{Ch2019,
  title = {Fermi-Arc-Induced Vortex Structure in Weyl Beam Shifts},
  author = {Chattopadhyay, Udvas and Shi, Li-kun and Zhang, Baile and Song, Justin C. W. and Chong, Y. D.},
  journal = {Phys. Rev. Lett.},
  volume = {122},
  issue = {6},
  pages = {066602},
  numpages = {6},
  year = {2019},
  month = {Feb},
  publisher = {American Physical Society},
  doi = {10.1103/PhysRevLett.122.066602},
  url = {https://link.aps.org/doi/10.1103/PhysRevLett.122.066602}
}

@article{weyl2,
  title = {Topological insulators and superconductors},
  author = {Qi, Xiao-Liang and Zhang, Shou-Cheng},
  journal = {Rev. Mod. Phys.},
  volume = {83},
  issue = {4},
  pages = {1057--1110},
  numpages = {0},
  year = {2011},
  month = {Oct},
  publisher = {American Physical Society},
  doi = {10.1103/RevModPhys.83.1057},
  url = {https://link.aps.org/doi/10.1103/RevModPhys.83.1057}
}

@article{weyl3,
  title = {Weyl and Dirac semimetals in three-dimensional solids},
  author = {Armitage, N. P. and Mele, E. J. and Vishwanath, Ashvin},
  journal = {Rev. Mod. Phys.},
  volume = {90},
  issue = {1},
  pages = {015001},
  numpages = {57},
  year = {2018},
  month = {Jan},
  publisher = {American Physical Society},
  doi = {10.1103/RevModPhys.90.015001},
  url = {https://link.aps.org/doi/10.1103/RevModPhys.90.015001}
}

@article{nl1,
  title = {Pseudospin Vortex Ring with a Nodal Line in Three Dimensions},
  author = {Lim, Lih-King and Moessner, Roderich},
  journal = {Phys. Rev. Lett.},
  volume = {118},
  issue = {1},
  pages = {016401},
  numpages = {6},
  year = {2017},
  month = {Jan},
  publisher = {American Physical Society},
  doi = {10.1103/PhysRevLett.118.016401},
  url = {https://link.aps.org/doi/10.1103/PhysRevLett.118.016401}
}

@article{nl2,
  title = {Conversion Rules for Weyl Points and Nodal Lines in Topological Media},
  author = {Sun, Xiao-Qi and Zhang, Shou-Cheng and Bzdu\ifmmode \check{s}\else \v{s}\fi{}ek, Tom\'a\ifmmode \check{s}\else \v{s}\fi{}},
  journal = {Phys. Rev. Lett.},
  volume = {121},
  issue = {10},
  pages = {106402},
  numpages = {7},
  year = {2018},
  month = {Sep},
  publisher = {American Physical Society},
  doi = {10.1103/PhysRevLett.121.106402},
  url = {https://link.aps.org/doi/10.1103/PhysRevLett.121.106402}
}

@article{Lilei2023,
  title = {Planar Hall effect in topological Weyl and nodal-line semimetals},
  author = {Li, Lei and Cao, Jin and Cui, Chaoxi and Yu, Zhi-Ming and Yao, Yugui},
  journal = {Phys. Rev. B},
  volume = {108},
  issue = {8},
  pages = {085120},
  numpages = {8},
  year = {2023},
  month = {Aug},
  publisher = {American Physical Society},
  doi = {10.1103/PhysRevB.108.085120},
  url = {https://link.aps.org/doi/10.1103/PhysRevB.108.085120}
}

@article{LvBQ2021,
  title = {Experimental perspective on three-dimensional topological semimetals},
  author = {Lv, B. Q. and Qian, T. and Ding, H.},
  journal = {Rev. Mod. Phys.},
  volume = {93},
  issue = {2},
  pages = {025002},
  numpages = {68},
  year = {2021},
  month = {Apr},
  publisher = {American Physical Society},
  doi = {10.1103/RevModPhys.93.025002},
  url = {https://link.aps.org/doi/10.1103/RevModPhys.93.025002}
}

@article{Li2024,
  title = {Anomalous shift in Andreev reflection from side incidence},
  author = {Li, Runze and Cui, Chaoxi and Liu, Ying and Yu, Zhi-Ming and Yang, Shengyuan A.},
  journal = {Phys. Rev. B},
  volume = {110},
  issue = {23},
  pages = {235114},
  numpages = {9},
  year = {2024},
  month = {Dec},
  publisher = {American Physical Society},
  doi = {10.1103/PhysRevB.110.235114},
  url = {https://link.aps.org/doi/10.1103/PhysRevB.110.235114}
}

@article{Yang2016,
author = {Yang, Shengyuan A.},
title = {Dirac and Weyl Materials: Fundamental Aspects and Some Spintronics Applications},
journal = {SPIN},
volume = {06},
number = {02},
pages = {1640003},
year = {2016},
doi = {10.1142/S2010324716400038},
URL = {https://doi.org/10.1142/S2010324716400038}
}

@article{Bansil2016,
  title = {Colloquium: Topological band theory},
  author = {Bansil, A. and Lin, Hsin and Das, Tanmoy},
  journal = {Rev. Mod. Phys.},
  volume = {88},
  issue = {2},
  pages = {021004},
  numpages = {37},
  year = {2016},
  month = {Jun},
  publisher = {American Physical Society},
  doi = {10.1103/RevModPhys.88.021004},
  url = {https://link.aps.org/doi/10.1103/RevModPhys.88.021004}
}

@article{Shuichi2007,
doi = {10.1088/1367-2630/9/9/356},
url = {https://dx.doi.org/10.1088/1367-2630/9/9/356},
year = {2007},
month = {sep},
publisher = {},
volume = {9},
number = {9},
pages = {356},
author = {Shuichi Murakami},
title = {Phase transition between the quantum spin Hall and insulator phases in 3D: emergence of a topological gapless phase},
journal = {New Journal of Physics}}

@article{Young2012,
  title = {Dirac Semimetal in Three Dimensions},
  author = {Young, S. M. and Zaheer, S. and Teo, J. C. Y. and Kane, C. L. and Mele, E. J. and Rappe, A. M.},
  journal = {Phys. Rev. Lett.},
  volume = {108},
  issue = {14},
  pages = {140405},
  numpages = {5},
  year = {2012},
  month = {Apr},
  publisher = {American Physical Society},
  doi = {10.1103/PhysRevLett.108.140405},
  url = {https://link.aps.org/doi/10.1103/PhysRevLett.108.140405}
}

@article{Wang2012,
  title = {Dirac semimetal and topological phase transitions in ${A}_{3}$Bi ($A=\text{Na}$, K, Rb)},
  author = {Wang, Zhijun and Sun, Yan and Chen, Xing-Qiu and Franchini, Cesare and Xu, Gang and Weng, Hongming and Dai, Xi and Fang, Zhong},
  journal = {Phys. Rev. B},
  volume = {85},
  issue = {19},
  pages = {195320},
  numpages = {5},
  year = {2012},
  month = {May},
  publisher = {American Physical Society},
  doi = {10.1103/PhysRevB.85.195320},
  url = {https://link.aps.org/doi/10.1103/PhysRevB.85.195320}
}

@article{YangSA2014,
  title = {Dirac and Weyl Superconductors in Three Dimensions},
  author = {Yang, Shengyuan A. and Pan, Hui and Zhang, Fan},
  journal = {Phys. Rev. Lett.},
  volume = {113},
  issue = {4},
  pages = {046401},
  numpages = {5},
  year = {2014},
  month = {Jul},
  publisher = {American Physical Society},
  doi = {10.1103/PhysRevLett.113.046401},
  url = {https://link.aps.org/doi/10.1103/PhysRevLett.113.046401}
}

@article{Mullen2015,
  title = {Line of Dirac Nodes in Hyperhoneycomb Lattices},
  author = {Mullen, Kieran and Uchoa, Bruno and Glatzhofer, Daniel T.},
  journal = {Phys. Rev. Lett.},
  volume = {115},
  issue = {2},
  pages = {026403},
  numpages = {5},
  year = {2015},
  month = {Jul},
  publisher = {American Physical Society},
  doi = {10.1103/PhysRevLett.115.026403},
  url = {https://link.aps.org/doi/10.1103/PhysRevLett.115.026403}
}

@article{chenY2015,
  title={Nanostructured carbon allotropes with Weyl-like loops and points},
  author={Chen, Yuanping and Xie, Yuee and Yang, Shengyuan A and Pan, Hui and Zhang, Fan and Cohen, Marvin L and Zhang, Shengbai},
  journal={Nano letters},
  volume={15},
  number={10},
  pages={6974--6978},
  year={2015},
  publisher={ACS Publications}
}

@article{YuR2015,
  title = {Topological Node-Line Semimetal and Dirac Semimetal State in Antiperovskite ${\mathrm{Cu}}_{3}\mathrm{PdN}$},
  author = {Yu, Rui and Weng, Hongming and Fang, Zhong and Dai, Xi and Hu, Xiao},
  journal = {Phys. Rev. Lett.},
  volume = {115},
  issue = {3},
  pages = {036807},
  numpages = {5},
  year = {2015},
  month = {Jul},
  publisher = {American Physical Society},
  doi = {10.1103/PhysRevLett.115.036807},
  url = {https://link.aps.org/doi/10.1103/PhysRevLett.115.036807}
}

@article{Kim2015,
  title = {Dirac Line Nodes in Inversion-Symmetric Crystals},
  author = {Kim, Youngkuk and Wieder, Benjamin J. and Kane, C. L. and Rappe, Andrew M.},
  journal = {Phys. Rev. Lett.},
  volume = {115},
  issue = {3},
  pages = {036806},
  numpages = {5},
  year = {2015},
  month = {Jul},
  publisher = {American Physical Society},
  doi = {10.1103/PhysRevLett.115.036806},
  url = {https://link.aps.org/doi/10.1103/PhysRevLett.115.036806}
}

@article{FangC2015,
  title = {Topological nodal line semimetals with and without spin-orbital coupling},
  author = {Fang, Chen and Chen, Yige and Kee, Hae-Young and Fu, Liang},
  journal = {Phys. Rev. B},
  volume = {92},
  issue = {8},
  pages = {081201},
  numpages = {5},
  year = {2015},
  month = {Aug},
  publisher = {American Physical Society},
  doi = {10.1103/PhysRevB.92.081201},
  url = {https://link.aps.org/doi/10.1103/PhysRevB.92.081201}
}

@article{zhong2016,
  title={Towards three-dimensional Weyl-surface semimetals in graphene networks},
  author={Zhong, Chengyong and Chen, Yuanping and Xie, Yuee and Yang, Shengyuan A and Cohen, Marvin L and Zhang, SB},
  journal={Nanoscale},
  volume={8},
  number={13},
  pages={7232--7239},
  year={2016},
  publisher={Royal Society of Chemistry}
}

@article{Liang2016,
  title = {Node-surface and node-line fermions from nonsymmorphic lattice symmetries},
  author = {Liang, Qi-Feng and Zhou, Jian and Yu, Rui and Wang, Zhi and Weng, Hongming},
  journal = {Phys. Rev. B},
  volume = {93},
  issue = {8},
  pages = {085427},
  numpages = {10},
  year = {2016},
  month = {Feb},
  publisher = {American Physical Society},
  doi = {10.1103/PhysRevB.93.085427},
  url = {https://link.aps.org/doi/10.1103/PhysRevB.93.085427}
}

@article{LiSi2018,
  title = {Almost ideal nodal-loop semimetal in monoclinic ${\mathbf{CuTeO}}_{3}$ material},
  author = {Li, Si and Liu, Ying and Fu, Botao and Yu, Zhi-Ming and Yang, Shengyuan A. and Yao, Yugui},
  journal = {Phys. Rev. B},
  volume = {97},
  issue = {24},
  pages = {245148},
  numpages = {8},
  year = {2018},
  month = {Jun},
  publisher = {American Physical Society},
  doi = {10.1103/PhysRevB.97.245148},
  url = {https://link.aps.org/doi/10.1103/PhysRevB.97.245148}
}

@article{bliokh2015,
  title={Spin--orbit interactions of light},
  author={Bliokh, Konstantin Yu and Rodr{\'\i}guez-Fortu{\~n}o, Francisco J and Nori, Franco and Zayats, Anatoly V},
  journal={Nature Photonics},
  volume={9},
  number={12},
  pages={796--808},
  year={2015},
  publisher={Nature Publishing Group}
}

@article{Fra1974,
  title = {Spatial displacement of electrons due to multiple total reflections},
  author = {Fradkin, D. M. and Kashuba, R. J.},
  journal = {Phys. Rev. D},
  volume = {9},
  issue = {10},
  pages = {2775--2788},
  numpages = {0},
  year = {1974},
  month = {May},
  publisher = {American Physical Society},
  doi = {10.1103/PhysRevD.9.2775},
  url = {https://link.aps.org/doi/10.1103/PhysRevD.9.2775}
}

@article{Sinitsyn2005,
  title = {Disorder effects in the anomalous Hall effect induced by Berry curvature},
  author = {Sinitsyn, N. A. and Niu, Qian and Sinova, Jairo and Nomura, Kentaro},
  journal = {Phys. Rev. B},
  volume = {72},
  issue = {4},
  pages = {045346},
  numpages = {11},
  year = {2005},
  month = {Jul},
  publisher = {American Physical Society},
  doi = {10.1103/PhysRevB.72.045346},
  url = {https://link.aps.org/doi/10.1103/PhysRevB.72.045346}
}

@article{Chen2008,
  title = {Tunable lateral displacement and spin beam splitter for ballistic electrons in two-dimensional magnetic-electric nanostructures},
  author = {Chen, Xi and Li, Chun-Fang and Ban, Yue},
  journal = {Phys. Rev. B},
  volume = {77},
  issue = {7},
  pages = {073307},
  numpages = {4},
  year = {2008},
  month = {Feb},
  publisher = {American Physical Society},
  doi = {10.1103/PhysRevB.77.073307},
  url = {https://link.aps.org/doi/10.1103/PhysRevB.77.073307}
}

@article{chen2011goos,
  title={Goos-H{\"a}nchen-like shifts for Dirac fermions in monolayer graphene barrier},
  author={Chen, Xi and Tao, J-W and Ban, Yue},
  journal={The European Physical Journal B},
  volume={79},
  number={2},
  pages={203--208},
  year={2011},
  publisher={Springer}}

@article{Sinitsyn2006,
  title = {Coordinate shift in the semiclassical Boltzmann equation and the anomalous Hall effect},
  author = {Sinitsyn, N. A. and Niu, Q. and MacDonald, A. H.},
  journal = {Phys. Rev. B},
  volume = {73},
  issue = {7},
  pages = {075318},
  numpages = {6},
  year = {2006},
  month = {Feb},
  publisher = {American Physical Society},
  doi = {10.1103/PhysRevB.73.075318},
  url = {https://link.aps.org/doi/10.1103/PhysRevB.73.075318}
}

@article{LiuGB2022,
  title = {Systematic investigation of emergent particles in type-III magnetic space groups},
  author = {Liu, Gui-Bin and Zhang, Zeying and Yu, Zhi-Ming and Yang, Shengyuan A. and Yao, Yugui},
  journal = {Phys. Rev. B},
  volume = {105},
  issue = {8},
  pages = {085117},
  numpages = {8},
  year = {2022},
  month = {Feb},
  publisher = {American Physical Society},
  doi = {10.1103/PhysRevB.105.085117},
  url = {https://link.aps.org/doi/10.1103/PhysRevB.105.085117}
}

@article{Zhang2022,
  title = {Encyclopedia of emergent particles in type-IV magnetic space groups},
  author = {Zhang, Zeying and Liu, Gui-Bin and Yu, Zhi-Ming and Yang, Shengyuan A. and Yao, Yugui},
  journal = {Phys. Rev. B},
  volume = {105},
  issue = {10},
  pages = {104426},
  numpages = {7},
  year = {2022},
  month = {Mar},
  publisher = {American Physical Society},
  doi = {10.1103/PhysRevB.105.104426},
  url = {https://link.aps.org/doi/10.1103/PhysRevB.105.104426}
}

@article{Zhang2023,
  title = {Encyclopedia of emergent particles in 528 magnetic layer groups and 394 magnetic rod groups},
  author = {Zhang, Zeying and Wu, Weikang and Liu, Gui-Bin and Yu, Zhi-Ming and Yang, Shengyuan A. and Yao, Yugui},
  journal = {Phys. Rev. B},
  volume = {107},
  issue = {7},
  pages = {075405},
  numpages = {7},
  year = {2023},
  month = {Feb},
  publisher = {American Physical Society},
  doi = {10.1103/PhysRevB.107.075405},
  url = {https://link.aps.org/doi/10.1103/PhysRevB.107.075405}
}

\end{document}